\documentclass[aps,prd,10pt,preprint,superscriptaddress,onecolumn,nofootinbib]{revtex4-1}

\usepackage{graphicx, epsfig} 
\usepackage{amsmath,amssymb,amsfonts,dsfont,mathrsfs,amsthm,mathtools,slashed}
\usepackage{bm} 
\usepackage{color}
\usepackage[usenames]{xcolor}
\usepackage{hyperref}
\usepackage{siunitx}
\hypersetup{colorlinks=true,urlcolor=blue,linkcolor=magenta,citecolor=blue,filecolor=blue}
\usepackage[normalem]{ulem}
\usepackage{array}
\usepackage{hyperref}
\usepackage{booktabs}
\usepackage{blindtext}
\usepackage{enumerate}

\usepackage[toc,page]{appendix}

\definecolor{armygreen}{rgb}{0.29, 0.33, 0.13}

\newcommand{\half}{\frac{1}{2}}

\newcommand{\JJ}{\mathds{J}}
\newcommand{\KK}{\mathds{K}}

\newcommand{\DD}{\mathds{D}}
\newcommand{\TT}{\mathds{T}}
\newcommand{\ZZ}{\mathds{Z}}

\newcommand{\QQ}{\mathds{Q}}
\newcommand{\QQb}{\overline{\mathds{Q}}}

\newcommand{\AAA}{\mathds{A}}
\newcommand{\FF}{\mathds{F}}
\newcommand{\EE}{\mathds{E}}
\newcommand{\OO}{\mathds{O}}

\newcommand{\calF}{\mathcal{F}}
\newcommand{\calX}{\mathcal{X}}
\newcommand{\ocalX}{\overline{\mathcal{X}}}
\newcommand{\calG}{\mathcal{G}}
\newcommand{\calH}{\mathcal{H}}

\newcommand{\calR}{\mathcal{R}}

\newcommand{\se}{\slashed{e}}

\newcommand{\psibar}{{\overline{\psi}}}

\newcommand{\gf}{\gamma_5 }

\newcommand*{\diag}{\operatorname{diag}}
\newcommand*{\Tr}{\operatorname{Tr}}

\newcommand{\diff}[1]{\text{d}#1}
\newcommand{\Diff}[1]{\text{D}#1}

\newcommand{\Lag}{\mathscr{L}}

\newcommand{\uantof}{Departamento de Física, Universidad de Antofagasta, Aptdo. 02800, Chile}

\newcommand{\cecs}{Centro de Estudios Cient\'{\i}ficos (CECs), Arturo Prat 514, Valdivia, Chile}

\newcommand{\icen}{Instituto de Ciencias Exactas y Naturales, Universidad Arturo Prat, Playa Brava 3256, 1111346, Iquique, Chile}

\newcommand{\fdec}{Facultad de Ciencias, Universidad Arturo Prat, Avenida Arturo Prat Chac\'on 2120, 1110939, Iquique, Chile}

\newcommand{\uss}{Universidad San Sebastián, General Lagos 1163, Valdivia, Chile}

\begin{document}

\title{Self duality in unconventional conformal supersymmetry}

\author{Pedro D. Alvarez}
\email{pedro.alvarez@uuss.cl}
\affiliation{\cecs}
\affiliation{\uss}
\affiliation{\uantof}

\author{Crist\'obal Corral}
\email{crcorral@unap.cl}
\affiliation{\icen}
\affiliation{\fdec}

\author{Jorge Zanelli}
\email{jorge.zanelli@uss.cl}
\affiliation{\cecs}
\affiliation{\uss}

\begin{abstract}

In this work, we study (anti-)self duality conditions in unconventional conformal supersymmetry. We focus on a theory constructed in a Townsend-MacDowell-Mansouri form for an $SU(2,2|N)$ gauge connection with matter fields in the adjoint representation. We find bosonic solutions that correspond to analytic gravitational instantons with nontrivial torsion. These configurations can be regarded as the torsional generalization of the Taub-NUT/Bolt-AdS and Eguchi-Hanson metric and they are (anti-)self-dual with respect to a generalized dual operator. We explore their global properties and show that they saturate a BPS bound.  
\end{abstract}

\maketitle


\section{Introduction}\label{intro} 

In recent years, a family of supersymmetric theories in which spin-1/2 fermions are combined with the bosonic non-Abelian gauge fields as part of a connection for a graded Lie algebra, has been explored~\cite{Alvarez:2011gd,Alvarez:2013tga,Alvarez:2021zsw}. In other words, bosons and fermions belong to the adjoint representation of a superalgebra rather than to the fundamental representation, as in the conventional form of supersymmetry. In this way, the resulting \emph{unconventional} supersymmetric (U-SUSY) theories do not exhibit matching of fermionic and bosonic degrees of freedom (no superpartners), the quantum numbers and couplings of all fields are those of standard QFT, and gravity is naturally built in the theory. Such models have a potential for a fermionic sector representing quarks and leptons and, in spite of being based on a local supersymmetric algebra, they contain no spin-3/2 gravitini.

In four dimensions, the resulting U-SUSY models can be seen as generalizations of Yang-Mills theory, where the dual operation (denoted by $\circledast$) allows writing a Lagrangian that is invariant under local Lorentz and internal gauge transformations, which are the expected unbroken symmetries of standard physics. The $\circledast$ operator can be found naturally in the superconformal algebra (see for instance \cite{Alvarez:2021zsw}). These algebras have been used to construct related theories of matter~\cite{Alvarez:2020qmy}, extended MacDowell-Mansouri supergravities~\cite{Alvarez:2021qbu} and a grand unified model~\cite{Alvarez:2022eew}. In these theories, the notion of duality for the bosonic fields is essentially the same as in Eq.~\eqref{YM-Hodge} for internal symmetry gauge fields and~\eqref{GR-Hodge} for the gravitational sector. In the case of fermions, what plays the role of the Levi-Civita symbol is the Clifford chirality operator $\gamma_5$ defined below. 

In Euclidean Yang-Mills (YM) theory, field configurations satisfying an (anti-)self duality condition, i.e. $F=\pm *F$,  correspond to absolute minima of the action.\footnote{Here, $F=\half F_{\mu \nu} \diff{x}^\mu \wedge \diff{x}^\nu$ is the YM field strength $2$-form, $*F=\half\varepsilon_{\mu \nu \lambda \rho}F^{\lambda \rho}\diff{x^\mu}\wedge\diff{x^\nu}$ its dual with and $\varepsilon_{\mu \nu \lambda \rho}$ the alternating Levi-Civita tensor in four-dimensions, and $\wedge$ denotes the wedge product of differential forms.} These \emph{instanton} configurations solve the field equations automatically by virtue of Bianchi identities, saturate a Bogomol'nyi–Prasad–Sommerfield (BPS) bound, and they describe the tunneling between topologically distinct vacua of the theory~\cite{Belavin:1975fg,tHooft:1976rip}. Instantons also play an important role in understanding non-perturbative aspects of QCD, such as the breaking of chiral symmetry~\cite{tHooft:1976snw,Diakonov:1985eg}.

Similarly, gravitational instantons are solutions to Einstein's equations of Euclidean general relativity satisfying (anti-)self-duality conditions, i.e. $R_{ab}=\pm \tfrac{1}{2}\epsilon_{abcd}R^{cd}\equiv\pm\tilde{R}_{ab}$, where $R^{ab}=\tfrac{1}{2}R^{ab}_{\ \mu\nu}\diff{x^\mu}\wedge\diff{x^\nu}$ is the Riemann curvature $2$-form and $\epsilon_{abcd}$ is the Levi-Civita symbol~\cite{Eguchi:1980jx}. Examples of gravitational instantons include the Euclidean Schwarzschild solution which describes a four-dimensional black hole, the Taub-NUT configuration \cite{Taub:1950ez,Newman:1963yy} and the Eguchi-Hanson instanton \cite{Eguchi:1978xp,Eguchi:1978gw} that describes an asymptotically locally Euclidean four-dimensional space whose radial slices are topologically $\mathbb{S}^3/\mathbb{Z}_2=\mathbb{RP}^3$. The interest in gravitational instantons was consolidated in the late 70s with the discovery of multi-center solutions \cite{Gibbons:1978tef} and the observation that those configurations are (hyper) K\"ahler geometries \cite{Hitchin:1979rts,ASENS_1979_4_12_2_269_0}.

The above examples highlight the importance of duality in the non-perturbative analysis of the topological sectors of the vacuum for the Euclidean continuation of a gauge theory. In each case, the notion of duality depends on the presence of a Hodge-like operator in the theory. For the Maxwell or Yang-Mills action, the relevant dual operation defines the kinetic term, that is,
\begin{equation} \label{YM-Hodge}
\frac{1}{2}\Tr[F\wedge *F] = -\frac{1}{4}\Tr[F_{\mu \nu}F^{\mu \nu}]\, \sqrt{|g|}\,\diff{}^4x  \,. 
\end{equation} 
In the gravitational case, on the other hand, the corresponding kinetic term reads
\begin{equation} \label{GR-Hodge}
 \frac{1}{2}\epsilon_{abcd}\, \mathscr{F}^{ab} \wedge \mathscr{F}^{cd} =
\mathscr{F}^{ab} \wedge \tilde{\mathscr{F}}_{ab} \,, 
\end{equation}
where the Yang-Mills field strength has been replaced by the (anti-)de Sitter curvature two-form, i.e. $\mathscr{F}^{ab}= R^{ab} \pm \lambda \, e^a \wedge e^b$, with $\lambda$ being a constant related to the (anti-)de Sitter radius and the Levi-Civita symbol in curved spacetime $\epsilon_{\mu \nu \lambda \rho}$ has been replaced by that in the tangent space, $\epsilon_{abcd}$~\cite{Townsend:1977xw,MacDowell:1977jt}. Additionally, $e^a=e^a_\mu\diff{x^\mu}$ denotes the vierbein field representing the metric structure of spacetime through the relation $g_{\mu\nu}=\eta_{ab}e^a_\mu e^b_\nu$ where $\eta_{ab}=\diag(-,+,+,+)$ is the Minkowski metric.\footnote{Here, we will work mostly in the Euclidean signature. Therefore, the relation between the metric and the vierbein reads $g_{\mu\nu}=\delta_{ab}e^a_\mu e^b_\nu$, where $\delta_{ab}=\diag(+,+,+,+)$ is the Euclidean metric.}

Then, a question that comes to mind is whether the notion of duality represented by $\circledast$ can also give rise to instanton-like configurations. In this paper, we explore the possibility afforded by this generalized duality operator to construct geometric instanton solutions in the matter-free sector.

The paper is organized as follows. In Sec.~\ref{sec:model}, we present the model we are interested in alongside their field equations and the consistent truncation to the bosonic sector. Section~\ref{sec:selfdual} is devoted to the (anti-)self duality condition in an unconventional conformal supersymmetric Townsend-MacDowell-Mansouri (TMM) theory for the $SU(2,2|N)$ group and to the analysis of its consequences for the action. In Sec.~\ref{sec:instantons}, we obtain (anti-)self-dual gravitational instantons that represent the torsional generalization of the Taub-NUT/Bolt-AdS and Eguchi-Hanson solutions in general relativity. Then, their global properties are analyzed in Sec.~\ref{sec:global} and our findings are summarized in Sec.~\ref{sec:summary}. Finally, Appendix~\ref{sec:representationApp} provides further details about the fundamental representation of $SU(2,2|N)$.

\section{Unconventional conformal supersymmetry}\label{sec:model} 

Let us illustrate the construction of a supersymmetric model in four dimensions with fermionic matter fields in the adjoint representation. Following Refs.~\cite{Alvarez:2013tga,Alvarez:2020qmy, Alvarez:2021qbu,Alvarez:2021zsw}, we start with a gauge connection for $SU(2,2|N)$. This algebra contains $\mathfrak{so}(3,2) \oplus \mathfrak{su}(N)$, representing local anti-de Sitter and internal symmetries in the bosonic sector, and supercharges in the fermionic sector. Then, as noticed by Townsend and by MacDowell and Mansouri, one is immediately faced with the fact that no dynamical gauge-invariant action for the entire $SU(2,2|N)$ group exists in four dimensions~\cite{Townsend:1977xw,MacDowell:1977jt}. This is due to the nonexistence of an $SU(2,2|N)$-invariant tensor in 4D. Not even a local $SO(3,2)$-invariant Lagrangian four-form can be found because there is no AdS-invariant fourth-rank tensor to contract the indices of the curvature forms. In fact, starting with the curvature two-form for the AdS connection, say $F_{AB}$, where uppercase Latin characters are AdS indices, the unique invariant four-form candidate Lagrangian in 4D would be of the form $\Lag = Q^{ABCD}F_{AB}\wedge F_{CD}$, with $Q^{ABCD}$ being a fourth-rank tensor, antisymmetric in $AB$ and $CD$. The tensor in question is of the form $Q^{ABCD}=\eta^{AC} \eta^{BD}-\eta^{AD} \eta^{BC}$, where $\eta^{AB}$ represents the inverse Cartan-Killing metric of $SO(3,2)$. In that case, however, the four-form Lagrangian $\Lag$ turns out to be a topological density: the Pontryagin form. So, the only way to obtain a meaningful action is by breaking the AdS symmetry down to the Lorentz group, the largest semisimple subgroup of $SO(3,2)$. 

The largest gauge symmetry of the action is a bosonic subgroup of $SU(2,2|N)$ which contains the Lorentz transformations and the internal $SU(N)$ gauge group. Therefore, due to the nonexistence of an $SU(2,2|N)$-invariant tensor, the introduction of the generalized dual operator $\circledast$ is crucial for constructing an action that is gauge-invariant for the subgroup. Indeed, an inspection of the gravity sector shows that an appropriate choice of the latter is such that $\langle \JJ_{ab} \circledast \JJ_{cd} \rangle \propto \epsilon_{abcd} $ so that local Lorentz symmetry is preserved \cite{MacDowell:1977jt}---see Eq.~\eqref{dualop} below for its definition. The $\mathfrak{su}(2,2|N)$ graded Lie algebra\footnote{For more details about the fundamental representation and algebra see Appendix~\ref{sec:representationApp}.} includes the generators of the conformal group, $\JJ_{a}, \KK_{a}, \JJ_{ab}$ and $\DD$, together with internal generators $\TT_I$ of $SU(N)$ and $\ZZ$ for $U(1)$, plus $N$ complex spinorial supercharges, $\QQ^\alpha_i$ and $\QQb^i_\alpha$. The $SU(2,2|N)$ connection can be written as
\begin{equation}\label{gaugeconnection}
\AAA= \Omega+\QQb_\alpha^i (\se \psi_i)^\alpha+(\overline{\psi}^i \se)_\alpha \QQ_i^\alpha\ \,,
\end{equation}
where we have denoted $(\se \psi_i)^\alpha \equiv (\gamma_a)^\alpha{}_\beta e^a \psi^\beta_i$, and the bosonic part of the connection is
\begin{equation}
\Omega= \half \omega^{ab}\JJ_{ab}+f^a \JJ_a+g^a\KK_a+h\DD+A^I \TT_I+A\ZZ \,.
\end{equation}
As usual, the field strength curvatures associated to this gauge connection are given by
\begin{align} 
 \FF =& \diff{}\AAA +\AAA \wedge \AAA 
 = \half \calF^{ab} \JJ_{ab}+\calF^a \JJ_a +\calG^a \KK_a +\calH \DD +\calF^I \TT_I +\calF \ZZ +\QQb \calX +\ocalX \QQ\,,
\end{align}
whose components along the $SU(2,2|N)$ generators are\footnote{Contracted spinor indexes will be omitted whenever unambiguous.}
    \begin{align}
\calF^{ab}&=\mathcal{R}^{ab}-\overline{\psi}^i\se\wedge\Sigma^{ab}\se\psi_i, & 
\calF^a&=\Diff{}f^a + g^a\wedge h+\frac{1}{2}\overline{\psi}^i\se\wedge\gamma^a\se\psi_i\,,  \\
\calG^a&=\Diff{}g^a+f^a\wedge h-\half \overline{\psi}^i\se\wedge\tilde{\gamma}^a\se\psi_i\,, &
\calH  &=\diff{}h+f^a\wedge g_a+\half \overline{\psi}^i\se\wedge\gf \se\psi_i\,, \\
\calF^I&= \diff{}A^I+\frac{1}{2} f_{JK}^{I}A^J\wedge A^K-i\overline{\psi}^i\se\wedge(\lambda^I)_i^{\ j}\se\psi_j\,, & 
\calF &=\diff{}A-\frac{i}{4 }\overline{\psi}^i\se\wedge\se\psi_i\,,
\end{align}
and the spinorial components of the field strength are
\begin{subequations}
    \begin{align}
\calX_i^\alpha &=\Diff{}(\se \psi_i)^\alpha+\frac{1}{2}f^a\wedge (\gamma_a\se \psi_i)^\alpha +\half g^a\wedge(\tilde{\gamma}_a\se \psi_i)^\alpha+\half h\wedge (\gf \se\psi_i)^\alpha\,, \\
\ocalX^i_\alpha&=-(\overline{\psi}^i\se)_\alpha\overleftarrow{\Diff{}}+\frac{1}{2} (\overline{\psi}^i\se\gamma_a)_\alpha\wedge f^a +\half (\overline{\psi}^i\se\tilde{\gamma}_a)_\alpha\wedge g^a+\half (\overline{\psi}^i\se\gf )_\alpha\wedge h\,.
\end{align}
\end{subequations}

Here, $\Diff{}$ is the exterior covariant derivative with respect to the Lorentz connection $\omega^{a}_{\ b}$ and $\diff{}$ is the exterior derivative. Moreover, $\gamma_a$ denotes the Dirac matrices satisfying the Clifford algebra $\left\{\gamma_a,\gamma_b\right\}=2\eta_{ab}\mathbb{I}$, the chiral $\gamma$-matrix is defined as $\gamma_5=i\gamma^0 \gamma^1 \gamma^2 \gamma^3$, and $\tilde{\gamma}^a=-\gamma_5\gamma^a$.  The generators of the Lorentz group in the spinorial representation are $\Sigma_{ab}=\frac{1}{4}\left[\gamma_a,\gamma_b\right]$ while the internal generators of the $SU(N)$ algebra, $\TT_I$, are anti-hermitian, i.e. $\TT_I^\dag=-\TT_I$ (also $\ZZ^\dag=-\ZZ$) and they satisfy the Lie algebra $[\TT_I,\TT_J]=f_{IJ}{}^K\TT_K$. The $\lambda$-matrices satisfy the commutation relation $[\lambda_I,\lambda_J]=f_{IJK} \lambda^K$, where indexes are raised/lowered with an Euclidean metric $\delta_{IJ}$. Additionally, we have defined $\mathcal{R}^{ab}=R^{ab}+f^a\wedge f^b-g^a\wedge g^b$ with $R^{ab}=\diff{}\omega^{ab}+\omega^a_{\ c}\wedge\omega^{cb}$ being the Lorentz curvature $2$-form. 

The generalized TMM action \cite{Townsend:1977xw,MacDowell:1977jt} for the $SU(2,2|N)$ group is,\footnote{For a similar construction see Refs.~\cite{Andrianopoli:2014aqa,Andrianopoli:2020zbl}}
\begin{equation}\label{action}
 I = - \int_{\mathcal{M}} \langle \FF\wedge \circledast \FF \rangle\,,
\end{equation}
where $\langle \ \cdot \ \rangle$ stands for an invariant supertrace and the dual operator $\circledast$ is defined through
\begin{equation}
\circledast \FF= S\left(\half \calR^{ab} \JJ_{ab}+F^a \JJ_a +G^a \KK_a\right)+(\varepsilon_1\ast)H \DD +(\varepsilon_2\ast) F^I \TT_I +(\varepsilon_3\ast)F \ZZ \,. \label{dualop}
\end{equation}
Here, $\varepsilon_i=\pm1$, $\forall i$, $\ast$ denotes the Hodge dual, and $S=i\gamma_5$ where $\gamma_5$ is the chiral gamma matrix defined such that $S^2=-\mathbb{I}$. More details of the generalized duality operator $\circledast$ can be found in Ref.~\cite{Alvarez:2021zsw}. The value of $\varepsilon_i$ can be fixed by demanding the absence of ghosts. The latter implies that $\varepsilon_1=\varepsilon_2=1$, while $\varepsilon_3=\pm1$ for $N<4$ and $N>4$, respectively. It is sufficient for the dual operator $\circledast$ to act linearly in the complexified algebra, that is parity odd and to satisfy $\circledast^2 = +1 (-1)$ in Euclidean (Lorentzian) signature, respectively. In fact, the conformal algebra contains a grading operator among the bosonic spacetime generators that allows for a natural definition of the dual operator (for more details see~\cite{Alvarez:2021zsw}). A grading among the bosonic generators of $\mathfrak{su}(2,2|N)$ can be defined by an operator $S$ such that
\begin{equation}\label{Sgrading}
 S \ G -\varepsilon G \ S = 0\,, \qquad \mathrm{where} \quad  \begin{cases}
\varepsilon = +1\,, \mathrm{if}\ G \in \{ \JJ_{ab}, \DD, \ZZ, \TT_I \} \,,\\
\varepsilon = -1\,, \mathrm{if}\ G \in \{ \JJ_a, \KK_a \} \,.
\end{cases}\,,
\end{equation}
define the $S$-even ($\varepsilon=1$) and $S$-odd ($\varepsilon=-1$) subspaces of the bosonic sector. The grading $S$ is independent of the grading that distinguishes fermionic and bosonic operators. A dual operator with properties mentioned above can be implemented in the spin 1/2 representation of conformal superalgebra $\mathfrak{su}(2,2|N)$ (see Appendix~\ref{sec:representationApp}). The spinor representation and the $S$-odd generators are essential to include matter fields in a geometric way \cite{Alvarez:2021zsw}. This structure can be used to define actions with a gauge symmetry $G^+ \subset SU(2,2)\times SU(N) \times U(1)$, where $G^+$ is the group generated by the $S$-even generators. With the appropriate definition of the dual operator $\circledast$, the following properties of super-traces involving products of $S$-even and $S$-odd operators are found
    \begin{align}\label{supertraces}
 &\langle \EE \circledast \OO \rangle = 0 = \langle \OO \circledast \EE \rangle\,, &
 &\langle \EE_1 \circledast \EE_2 \rangle=\langle \EE_2 \circledast \EE_1 \rangle\,,&
 &\langle \OO_1 \circledast \OO_2 \rangle=-\langle \OO_2 \circledast \OO_1 \rangle\,,
\end{align}
which can be understood as a $\circledast$-grading of the invariant super-traces of the superalgebra, where $\EE_i$ and $\OO_i$ represent two-form generators that are even and odd in this grading. The set of $\circledast$-even generators contains the $S$-even bosonic generators and the supercharges. In particular, Eq.~\eqref{supertraces} implies that
\begin{equation}\label{proptrace1}
 \langle \FF^- \wedge\circledast \FF^- \rangle = 0\,,
\end{equation}
where $\FF^-$ are components of the curvature along the $S$-odd generators, namely,
\begin{equation}
\FF^-=\calF^a \JJ_a +\calG^a \KK_a\,. 
\end{equation}
In this way, the compensating fields $f^a$ and $g^a$ do not acquire independent kinetic terms in the action principle \eqref{action}. Thus, variations with respect to $f^a$ and $g^a$ yield algebraic equations for them, which means that they are actually auxiliary fields.

The field equations of the theory are obtained by demanding arbitrary variations of the action \eqref{action} with respect to the gauge connection, giving
\begin{equation}\label{fieldeqs}
 D_\AAA\circledast (\FF-\FF^-)=0\,,
\end{equation}
where $D_\AAA$ stands for the covariant derivative with respect to the gauge connection $\AAA$, provided the supertrace is nondegenerate [cf. Eqs.~\eqref{trace1}-\eqref{trace-1}]. In a purely bosonic background, the field equations take the simpler form $D_\Omega\circledast \FF^+=0$, and in the sector without internal gauge fields, $A^I = 0 = A$, the field equations, expanded along the $SU(2,2|N)$ generators, take the form
\begin{subequations}
    \begin{align}
 \Diff{} \calR^{cd} &=0\,, \label{topeq}\\ 
 \diff{}\ast (H+ f^b\wedge g_b)&=0\,,\\
 g_a\wedge\ast (H+ f^b\wedge g_b)-\half \epsilon_{abcd}f^b\wedge \calR^{cd}&=0\,, \\
 f_a\wedge\ast (H+ f^b\wedge g_b)-\half \epsilon_{abcd}g^b\wedge \calR^{cd}&=0\,,
\end{align}
\end{subequations}
where $H := dh$. Now, the same field equations can be obtained varying with respect to $\omega^{ab}$, $h$, $f^a$, and $g^a$ the truncated action
\begin{equation}\label{actionexpandbos}
\Lag_{\rm bos}=\frac{1}{4} \epsilon_{abcd}\calR^{ab}\wedge\calR^{cd}- (H+f^a\wedge g_a)\wedge\ast (H+f^b\wedge g_b)\,.
\end{equation}

The fields $f^a$ and $g^a$, associated with translations and special conformal transformations, are auxiliary (they satisfy algebraic equations). Since the gauge invariance corresponding to their associated transformations is broken, their equations can be algebraically solved in a satisfactory manner, provided they do not break the remaining gauge symmetries. There is a particular choice for these fields that respects local Lorentz invariance, that is, $f^a = \rho(x) e^a$, $g^a = \sigma(x) e^a$, where $\rho(x)$ and $\sigma(x)$ are arbitrary scalar functions. Consistent with this assumption $H$ must vanish, and therefore one can take $h=0$. Then, we get $\mathcal{R}^{ab} = R^{ab} - \Lambda(x) e^a\wedge e^b$, where  $\Lambda(x):=\sigma^2(x)-\rho^2(x)$. With these definitions, the field equations of the system are
\begin{subequations}
    \begin{align}\label{eomw}
 \Diff{} \calR^{cd}&=0\,, \\  \epsilon_{abcd}\calR^{bc}\wedge e^d&=0\,.\label{eomfog}
\end{align}
\end{subequations}
Equation~\eqref{eomw} resembles the Bianchi identity but its meaning is different because $\calR$ is not the curvature for the Lorentz connection.\footnote{Although the (anti-)self-duality condition requires Euclidean signature, we continue to refer to the spacetime rotations as Lorentz transformations or $SO(4)$ rotations indistinctly.} Indeed, this equation reduces to
\begin{align}\label{eqforLambda}
\diff{\ln\Lambda(x)} e_a\wedge  e_b = -T_{[a}\wedge e_{b]}\,,
\end{align}
which is an algebraic equation for the torsion that can be solved in terms of derivatives of $\Lambda(x)$. Equation ~\eqref{eomfog}, on the other hand, is basically Einstein's equations with variable $\Lambda$. In what follows we will consider solutions of these equations that satisfy (anti-)self duality conditions.

\section{(anti-)Self-dual configurations}\label{sec:selfdual} 

Since we are interested in (anti-)self-dual configurations, we focus on the Euclidean section of the theory by performing the analytic continuation of the Lorentzian time into the Euclidean one. This leads to an Euclidean superconformal group, whose properties have been discussed in Refs.~\cite{Lukierski:1983jg,Lukierski:1983qc,Lukierski:1986jw}. The existence of the generalized dual operator $\circledast$ provides us with a concrete notion of (anti-)self-duality, that is,
\begin{equation}\label{selfdual}
 \circledast (\FF-\FF^-) = \pm (\FF-\FF^-)\,.
\end{equation}
This condition is motivated by the definition of the operator $\circledast$ and the form of the field equations~\eqref{fieldeqs}. As with the standard notion of duality mentioned above, the $\circledast$-graded traces \eqref{supertraces} imply that a (anti-)self-dual configurations \eqref{selfdual} solve the field equations \eqref{fieldeqs}. These configurations saturate the BPS bound as well, that is,
\begin{equation}
  \int_{\mathcal{M}} \langle \FF \wedge \circledast \FF \rangle = \pm\int_{\mathcal{M}} \langle \FF \wedge \FF \rangle = \mp 8 \pi^2 C\,,  
\end{equation}
where $C$ denotes the Chern-Pontryagin index for Euclidean $SU(2,2|N)$. Indeed, one can supplement the action~\eqref{action} with a topological invariant of the Pontryagin class without modifying the field equations, in order to set all the (anti-)self-dual configurations as ground states of the theory (see for instance Refs.~\cite{Miskovic:2009bm,Araneda:2016iiy}), namely, 
\begin{align}
  I \mp\int_{\mathcal{M}}\langle \FF \wedge \FF \rangle = \mp\frac{1}{2}\int_{\mathcal{M}}\langle\left(\FF \pm \circledast \FF \right)\wedge\circledast\left(\FF \pm \circledast \FF \right)\rangle\,.
\end{align}
One can check directly that this action vanishes for configurations satisfying \eqref{selfdual} by virtue of ~\eqref{proptrace1}, extending the moduli space of vacuum configurations to all (anti-)self-dual states. 

We consider the matter-free (anti-)self-dual configurations, that is, with no fermions and vanishing internal gauge fields: $\psi = 0$, $\psibar = 0$, $A^I = 0$, $A = 0$. We take the fields $f^a = \rho(x) e^a$, $g^a = \sigma(x) e^a$, and $h=0$, then condition~\eqref{selfdual} alongside the relevant components of the curvature gives
\begin{align}
\calR_{ab} &=\pm\half \epsilon_{abcd}\calR^{cd}\equiv\pm\tilde{\calR}_{ab}\,.\label{selfdualRcal}
\end{align}
Combining this condition with \eqref{eomfog} gives $R^a{}_b e^b=0$, which implies covariantly constant Torsion, $\Diff{} T^a=0$. This, in turn, forces $\tilde{\calR}^{ab} = \ast \mathcal{R}^{ab}$, which relates two different notions of (anti-)self duality on-shell.

Therefore, if Eqs.~\eqref{selfdualRcal} and~\eqref{eqforLambda} are satisfied, the field equations are solved automatically due to (anti-)self duality. Moreover, one can show that the Euclidean on-shell action is proportional to the topological Chern-Pontryagin index when evaluated on (anti-)self-dual configurations satisfying Eq.~\eqref{selfdualRcal}, since the Lagrangian~\eqref{actionexpandbos} becomes
\begin{align}
    \Lag_{\rm bos} = \frac{1}{4} \epsilon_{abcd}\calR^{ab}\wedge\calR^{cd} =  \pm\half \calR_{ab}\wedge\calR^{ab} = \pm\half R_{ab}\wedge R^{ab}\,,
\end{align}
where the last equality is a consequence of the Bianchi identity $\Diff{T^a} = R^{a}_{\ b}\wedge e^b$ and the condition $\Diff{T^a}=0$. Hence, the Euclidean on-shell action is
\begin{align}\label{IequalP1}
I=\int_{\mathcal{M}}\Lag_{\rm bos} = \pm\frac{1}{2}\int_{\mathcal{M}}R_{ab}\wedge R^{ab} =\pm4\pi^2 P_1[\mathcal{M}]
\end{align}
where $P_1[\mathcal{M}]$ is the $SO(4)$ Chern-Pontryagin index that measures the differences between the number of harmonic self-dual and anti-self-dual forms on $\mathcal{M}$. 

In what follows, we present a family of gravitational instantons with nontrivial torsion in unconventional conformal supersymmetric TMM theory; some of them being (anti-)self-dual in the sense of Eq.~\eqref{selfdualRcal}. They represent concrete solutions that exhibit the conditions discussed here.

\section{Gravitational instantons}\label{sec:instantons} 

In order to find (anti-)self-dual solutions to the field equations~\eqref{eomw} and~\eqref{eomfog}, we construct an ansatz for the vierbein, the ``Lorentz" $SO(4)$ connection and $\Lambda(x)$, based on the $U(1)$-fibration over K\"ahler manifolds. To this end, we consider the left-invariant Maurer-Cartan forms of $SU(2)$, $\sigma_i$ with $i=1,2,3$; they are
\begin{subequations}
    \begin{align}
\sigma_1 &= \cos\tau\,\diff{\vartheta} + \sin\tau\sin\vartheta\,\diff{\varphi}\,,\\
\sigma_2 &= -\sin\tau\,\diff{\vartheta} + \cos\tau\sin\vartheta\,\diff{\varphi}\,, \\
\sigma_3 &= \diff{\tau} + \cos\vartheta\,\diff{\varphi}\,.
\end{align}
\end{subequations}
These $\mathfrak{su}(2)$-valued $1$-forms satisfy the condition $\diff{\sigma_i}=-\tfrac{1}{2}\epsilon_{ijk}\sigma^j\wedge\sigma^k$. Using them as building blocks, we consider a metric ansatz whose radial slices are conformally related to Berger's sphere, that is,
\begin{align}\label{metricansatz}
\diff{s^2} &= \frac{\diff{r^2}}{f(r)} + f(r)J^2(r)\sigma_3^2 + h(r)\left(\sigma_1^2+\sigma_2^2 \right)\,,
\end{align}
where $\sigma_1^2+\sigma_2^2 = \diff{\vartheta^2}+\sin^2\vartheta\diff{\varphi^2}$ is the line element of the unit $2$-sphere. The metric~\eqref{metricansatz} is locally invariant under the action of the isometry group $SU(2)\times U(1)$ and possesses the relevant properties for analyzing different (anti-)self-dual configurations. Additionally, using the relation $g_{\mu\nu} = \delta_{ab} e^a_\mu e^b_\nu$ in Euclidean signature, we choose a vierbein basis as
\begin{align}\label{eansatz}
    e^1 &= \sqrt{h(r)}\,\sigma_1\,, & e^2 &= \sqrt{h(r)}\,\sigma_2\,, & e^3 &= \sqrt{f(r)}\,J(r)\,\sigma_3\,, & e^4 &= \frac{\diff{r}}{\sqrt{f(r)}}\,.
\end{align}
Analogously, demanding the Lorentz connection to remain invariant under the action of the isometry group of the metric, the connection satisfies
\begin{subequations}\label{wansatz}
\begin{align}
    \omega^{12} &= \omega_1(r)\,\sigma_3 + \omega_2(r)\,e^4 + \mathring{\omega}^{12}\,, & \omega^{13} &= \omega_3(r)\,\sigma_1 - \omega_4(r)\,\sigma_2\,, \\ \omega^{14} &= \omega_5(r)\,\sigma_1 - \omega_6(r)\,\sigma_2\,, & \omega^{23} &= \omega_4(r)\,\sigma_1 + \omega_3(r)\,\sigma_2\,, \\ \omega^{24} &= \omega_6(r)\,\sigma_1 + \omega_5(r)\,\sigma_2\,, & \omega^{34} &= \omega_7(r)\,\sigma_3 + \omega_8(r)\,e^4\,,
\end{align}
\end{subequations}
where $\omega_i(r)$, with $i=1,...,8$, are arbitrary functions of the radial coordinate only, to be determined from the field equations, and $\mathring{\omega}^{ab}$ is the torsion-free Levi-Civita connection defined by $\diff{e^a}+\mathring{\omega}^{a}_{\ b}\wedge e^b=0$. Equivalently, the most general field $\Lambda(x)$ compatible with the action of $SU(2)\times U(1)$ isometry group implies that $\Lambda=\Lambda(r)$. Then, the field equations for the connection~\eqref{eomw} are solved by
\begin{subequations}\label{eomfogsolutions}
    \begin{align}
        \omega_1 &= \omega_2 = \omega_3 = \omega_6 = \omega_8 = 0\,, &
        \omega_4 &= -\frac{J}{2}\sqrt{\frac{f}{h}}\,, \\
        \omega_5 &= \frac{\left(h\Lambda \right)'}{2\Lambda}\sqrt{\frac{f}{h}}\,,&
        \omega_7 &= \frac{\left(f\Lambda J^2\right)'}{2\Lambda J}\,,
    \end{align}
\end{subequations}
where $f=f(r)$, $h=h(r)$, $J=J(r)$, $\Lambda=\Lambda(r)$, and prime denotes differentiation with respect to the radial coordinate $r$. 

The form of these ans\"atze, together with the solution of the equation for the connection, allow the construction of two interesting families of gravitational instantons of the Taub-NUT and Eguchi-Hanson types possessing nontrivial torsion.

\subsection{Taub-NUT/Bolt-AdS-like solutions} 

The Taub-NUT metric, first found as a one-parameter extension of the Schwarzschild solution in vacuum general relativity~\cite{Taub:1950ez,Newman:1963yy}, is usually interpreted as a gravitational dyon in the sense that the new parameter --the NUT charge-- introduces a gravitomagnetic interaction between the background geometry and test particles following geodesics~\cite{Lynden-Bell:1996dpw}. The solution presents a topological defect --the Misner string~\cite{Misner:1963fr}--, which is the gravitational analog of the Dirac string in the magnetic monopole. The string can be rendered unobservable by performing an improper coordinate transformation that, in the Lorentzian signature, introduces closed time-like curves~\cite{Misner:1963fr}. Thus, the Taub-NUT metric is usually studied in the Euclidean continuation, which bears a close resemblance to instantons in Yang-Mills theory~\cite{Hawking:1976jb,Eguchi:1977iu,Gibbons:1979xm,Eguchi:1980jx}. However, recent developments have led to consider the Lorentzian Taub-NUT metric again since it was shown that the Misner string is transparent to geodesic observers~\cite{Clement:2015cxa}, allowing an interpretation of the NUT charge as a genuine thermodynamic variable~\cite{Hennigar:2019ive,Bordo:2019slw,Bordo:2019tyh,BallonBordo:2019vrn,Durka:2019ajz,Wu:2019pzr,Chen:2019uhp,BallonBordo:2020mcs}.  

The Taub-NUT solution with nontrivial torsion in the unconventional conformal supersymmetric model studied here is obtained by replacing the solutions \eqref{eomfogsolutions} into ~\eqref{eomfog} alongside the radial-coordinate choice $h(r)=r^2-n^2$, with $n$ being the NUT charge. Thus, $h(r)$ represents the area coordinate with the origin at $r=n$. The hypersurfaces of constant $r$ are conformally the Berger's sphere. In general, $n\in\mathbb{R}$. However, in this work, we shall focus on the case $n\in\mathbb{R}^+$.

The remaining functions that solve the field equations are found to be $J(r)=2n$ and
\begin{subequations}\label{TaubNUTsol}
    \begin{align}
    f(r) &= \frac{a\left[r\,b^2 + 3r^2(r+b) + bn^2  \right]}{b\left(r^2-n^2 \right)} - \frac{r(r+b)^2}{b\left(r^2-n^2 \right)} + \frac{(r+b)^3\left(r\,b+n^2\right)}{b\ell^2\left(r^2-n^2\right)}\,, \label{fsolgen}\\
    \Lambda(r) &= \frac{a}{(r+b)^2}\,,
\end{align}
\end{subequations}
where $a$, $b$ and $\ell$ are integration constants. This is a four-parameter family of solutions with nontrivial torsion sourced by $\Lambda(x)$ [see Eq.~\eqref{eqforLambda}]. Explicitly, the components of torsion associated to the solution in Eq.~\eqref{TaubNUTsol} can be expressed as
\begin{align}
    T^a = -\frac{\sqrt{f(r)}}{r+b}\; e^a\wedge e^4 \,,
\end{align}
which vanishes at the zeroes of the metric function $f(r)$; the nuts and bolts cases are discussed below. On the other hand, the components of the Lorentz curvature $2$-form for this solution are 
\begin{subequations}\label{RcompTN}
    \begin{align}
        R^{12} &=\mathscr{A}(r)\,e^1\wedge e^2 - 2\mathscr{B}(r)\,e^3\wedge e^4\,, & R^{13} &= \mathscr{C}(r)\, e^1\wedge e^3 - \mathscr{B}(r)\,e^2\wedge e^4\,, \\ 
        R^{14} &= \mathscr{C}(r)\,e^1\wedge e^4 + \mathscr{B}(r)e^2\wedge e^3\,, & R^{23} &= \mathscr{C}(r)\,e^2\wedge e^3 + \mathscr{B}(r)\,e^1\wedge e^4\,,\\
        R^{24} &= \mathscr{C}(r)\,e^2\wedge e^4 - \mathscr{B}(r)\,e^1\wedge e^3\,, & R^{34} &= \mathscr{D}(r)\,e^3\wedge e^4 - 2\mathscr{B}(r)\,e^1\wedge e^2\,,
    \end{align}
    \end{subequations}
where
\begin{subequations}
    \begin{align}
        \mathscr{A}(r) &= \frac{1}{r^2-n^2}\left[1 - \frac{f\left(b^2r^2+3n^2r^2+8bn^2r+3b^2n^2+n^4\right)}{\left(r^2-n^2\right)(r+b)^2} \right] \,, \\
        \mathscr{B}(r) &= \left[\frac{fn}{2\left(r^2-n^2\right)} \right]' \,, \\
        \mathscr{C}(r) &= -\left[\frac{f\left(br+n^2 \right)}{2\left(r^2-n^2 \right)(r+b)} \right]' - \frac{f\left(b^2-n^2 \right)}{2\left(r^2-n^2 \right)(r+b)^2} \,, \\
        \mathscr{D}(r) &= -\frac{f''}{2} + \frac{f'}{r+b} - \frac{f}{(r+b)^2}\,.
    \end{align}
\end{subequations}

Additionally, the asymptotic behavior of the solution as $r\to\infty$ is given by
\begin{subequations}
    \begin{align}\notag
    f(r) &\simeq \frac{r^2}{\ell^2} + \frac{r}{b\ell^2}\left(3a\ell^2-\ell^2+n^2+3b^2\right)\ +  \frac{1}{\ell^2}\left(3a\ell^2-2\ell^2+4n^2+3b^2\right) \\ &+ \frac{1}{r\,b\ell^2}\left[\ell^2\left(3an^2+(a-1)b^2-n^2 \right)+b^4+n^4+6b^2n^2 \right] + \mathcal{O}(r^{-2})\,, \\
    \Lambda(r) &\simeq \frac{a}{r^2} - \frac{2ab}{r^3} + \frac{3ab^2}{r^4} + \mathcal{O}(r^{-5})\,.
\end{align}
\end{subequations}
The metric function exhibits the typical linear behavior in the radial coordinate as $r\to\infty$ that appears in solutions to conformal gravity~\cite{Riegert:1984zz,Lu:2012xu,Liu:2013fna,Corral:2021xsu}, producing a space with weakened local AdS asymptotics. However, as shown by Maldacena in Ref.~\cite{Maldacena:2011mk}, one can impose Newmann boundary conditions on the Fefferman-Graham expansion to select only the Einstein modes of the theory, reducing four-dimensional conformal gravity to Einstein-AdS gravity (see also Refs.~\cite{Anastasiou:2016jix,Anastasiou:2020mik}). In the case of the torsional Taub-NUT instanton presented here, the Riemannian curvature, i.e. the one constructed from the Levi-Civita connection $\mathring{\omega}^{a}_{\ b}$, possesses an asymptotically locally AdS behavior with curvature radius $\ell^{-2}$. The Lorentz curvature $2$-form is asymptotically locally flat rather than asymptotically locally AdS. This is due to the fact that the leading order in the asymptotic expansion of the torsion, i.e.,
\begin{align}
    T^a = -\left[ \frac{1}{\ell} + \frac{3a\ell^2+b^2-\ell^2+n^2}{2b\ell\,r} + \mathcal{O}(r^{-2}) \right]e^a\wedge e^4\,,
\end{align}
is asymptotically constant. This constant in the contorsion contribution to the Lorentz curvature cancels the constant term from the Riemann curvature. Then, the fall-off of the Lorentz curvature $2$-form in Eq.~\eqref{RcompTN} is encoded in the asymptotic expansion of its components as $r\to\infty$, that is,
\begin{subequations}
    \begin{align}
    \mathscr{A}(r) &= \frac{\ell^2-b^2-3n^2}{\ell^2\,r^2} + \mathcal{O}(r^{-3})\,, \\ 
    \mathscr{B}(r) &= \frac{n\left(\ell^2-n^2-3a\ell^2-3b^2 \right)}{2b\ell^2 r^2} + \mathcal{O}(r^{-3})\,,  \\ 
    \mathscr{C}(r) &= \frac{3a\ell^2+b^2-\ell^2+3n^2}{2\ell r^2} + \mathcal{O}(r^{-3}) \,,\\ 
    \mathscr{D}(r) &= \frac{\ell^2-b^2-3n^2}{\ell^2 r^2} + \mathcal{O}(r^{-3})\,.
\end{align}
\end{subequations}
These configurations are therefore asymptotically locally Lorentz-flat solutions. Similar locally Lorentz-flat configurations are solutions of three-dimensional Einstein-AdS gravity~\cite{Alvarez:2014uda}, which can also be formulated as a Chern-Simons theory for the AdS group (for a review see~\cite{Zanelli:2005sa,Hassaine:2016amq}). On the other hand, asymptotically locally Lorentz-flat black holes have been reported in Ref.~\cite{Alvarez:2022wcj} in four-dimensional unconventional conformal supersymmetric TMM theory. 

There are two distinct topologies that can be obtained from the solution ~\eqref{TaubNUTsol}. They are characterized by the dimensionality of the set of fixed points of the Killing vector $\xi=\partial_\tau$. In particular, they correspond to nuts or bolts if either of the following conditions are met
\begin{align}\label{nut}
&\text{Nut:}& f(n) &= 0\, & &\mbox{and}& \beta_\tau &= \frac{4\pi}{f'(r)}\bigg|_{r=n}\,, \\
\label{bolt}
&\text{Bolt:}& f(r_h) &= 0\, & &\mbox{and}& \beta_\tau &= \frac{4\pi}{f'(r)}\bigg|_{r=r_h}\,,
\end{align}
where $\beta_\tau$ is the period of the Euclidean time, i.e. $\tau\sim\tau+\beta_\tau$ and $r_h>n$ is the bolt radius. The NUT condition ~\eqref{nut} fixes one integration constant as
\begin{align}\label{anut}
    a = \frac{(b+n)[\ell^2-(b+n)^2]}{\ell^2(b+3n)}\equiv a_{\rm nut}\,.
\end{align}
Then, the metric function $f(r)$ and the scalar $\Lambda(r)$ become
\begin{subequations}\label{solnut}
    \begin{align}
f_n(r) &= \frac{(r-n)(2r+b+n)}{(b+3n)(r+n)}+ \frac{(r-n)^2\left[r\,b+3n(r+b)+n^2\right]}{\ell^2(b+3n)(r+n)}\,, \\
\Lambda_n(r) &= \frac{(b+n)\left[\ell^2-(b+n)^2 \right]}{\ell^2(b+3n)(r+b)^2}\,.
\end{align}
\end{subequations}
Remarkably, this solution is locally self-dual in the sense of ~\eqref{selfdualRcal} and possesses nontrivial torsion sourced by $\Lambda(x)$. The absence of conical singularities at $r=n$ implies that the Euclidean-time coordinate has to be identified as $0\leq\tau\leq8\pi n$, rendering the topology near the origin that of $\mathbb{R}^4$. 

\begin{figure}[h]
  \centering
  \includegraphics[scale=0.25]{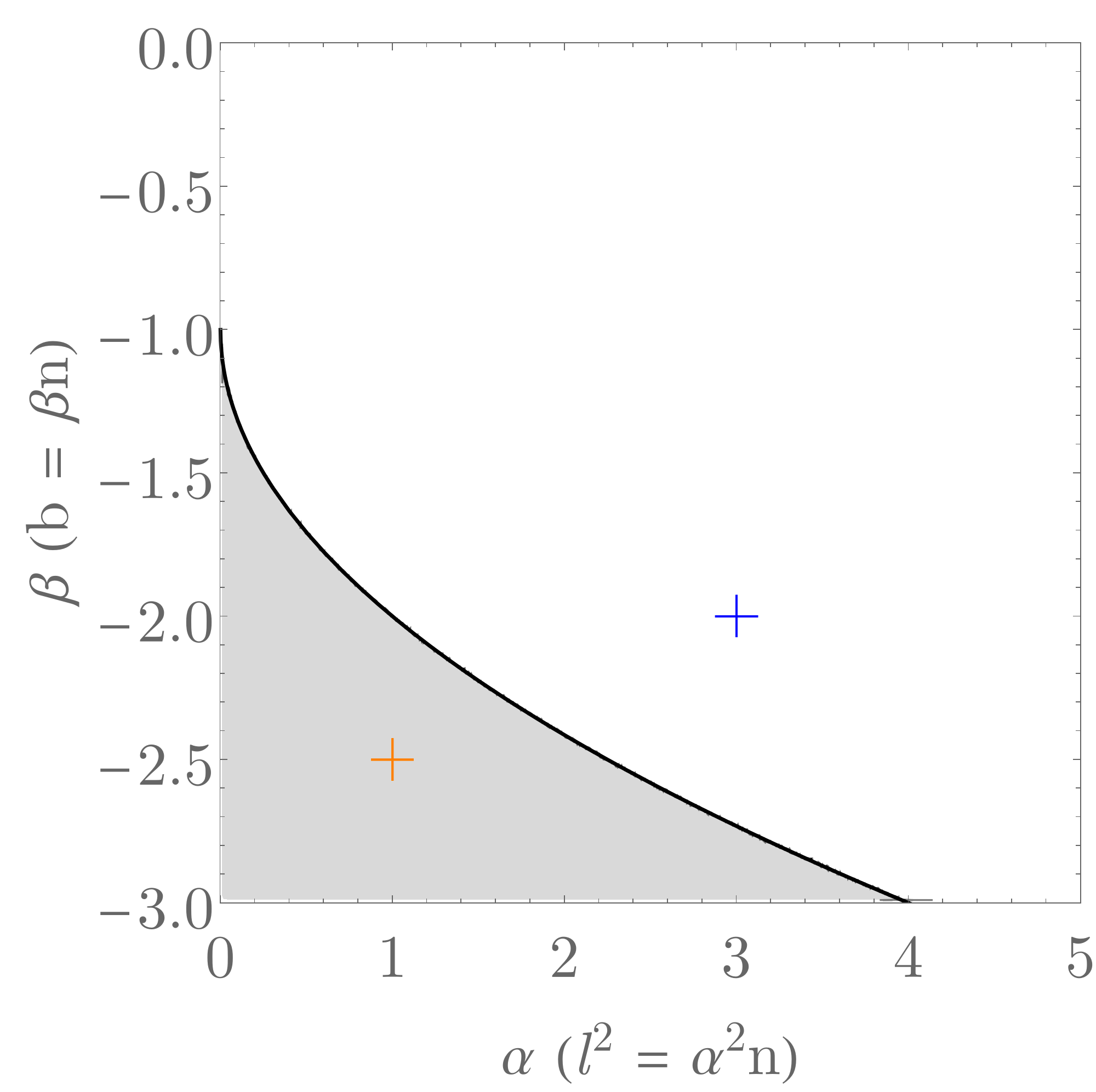}
  \includegraphics[scale=0.25]{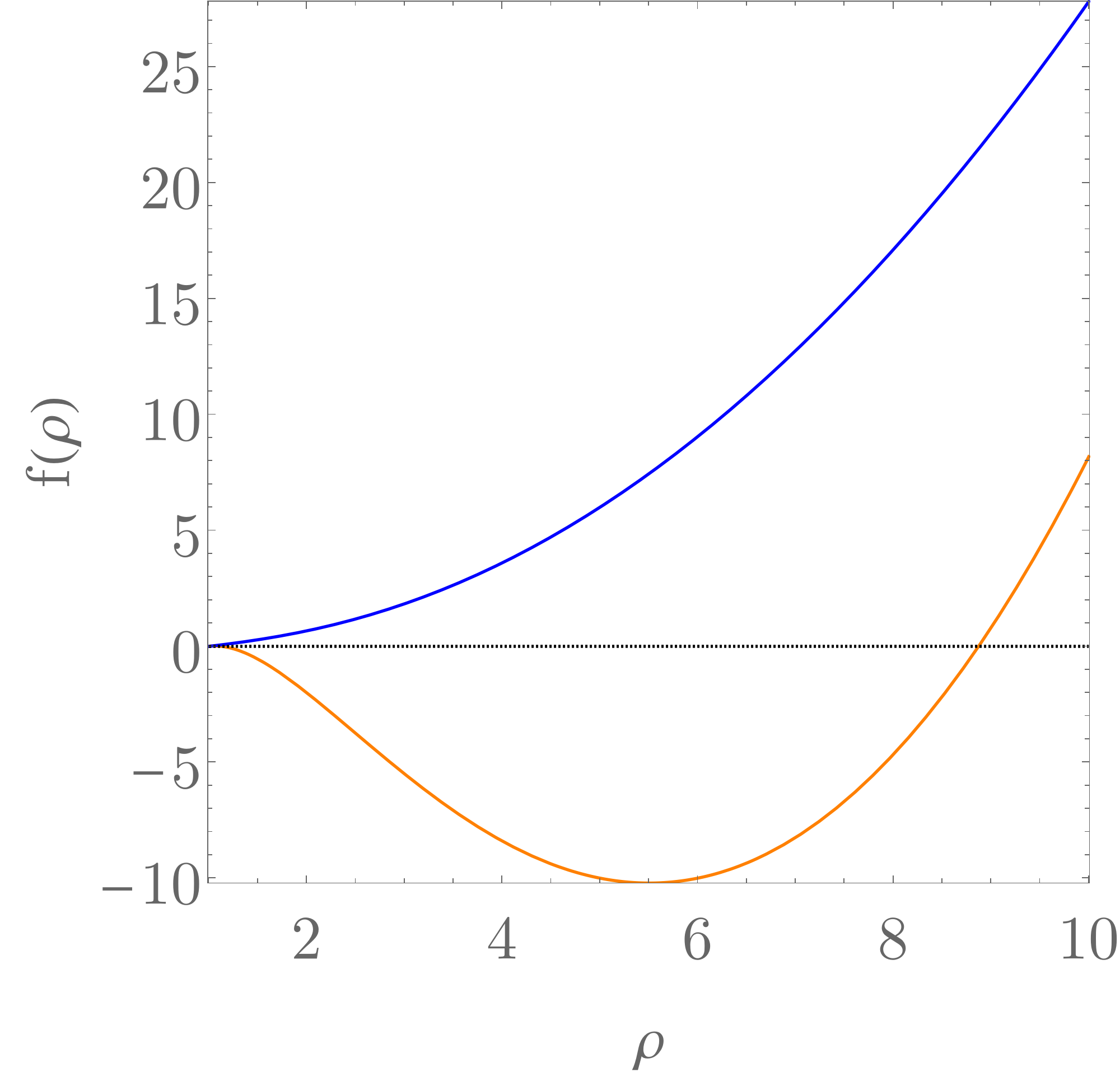}
  \includegraphics[scale=0.25]{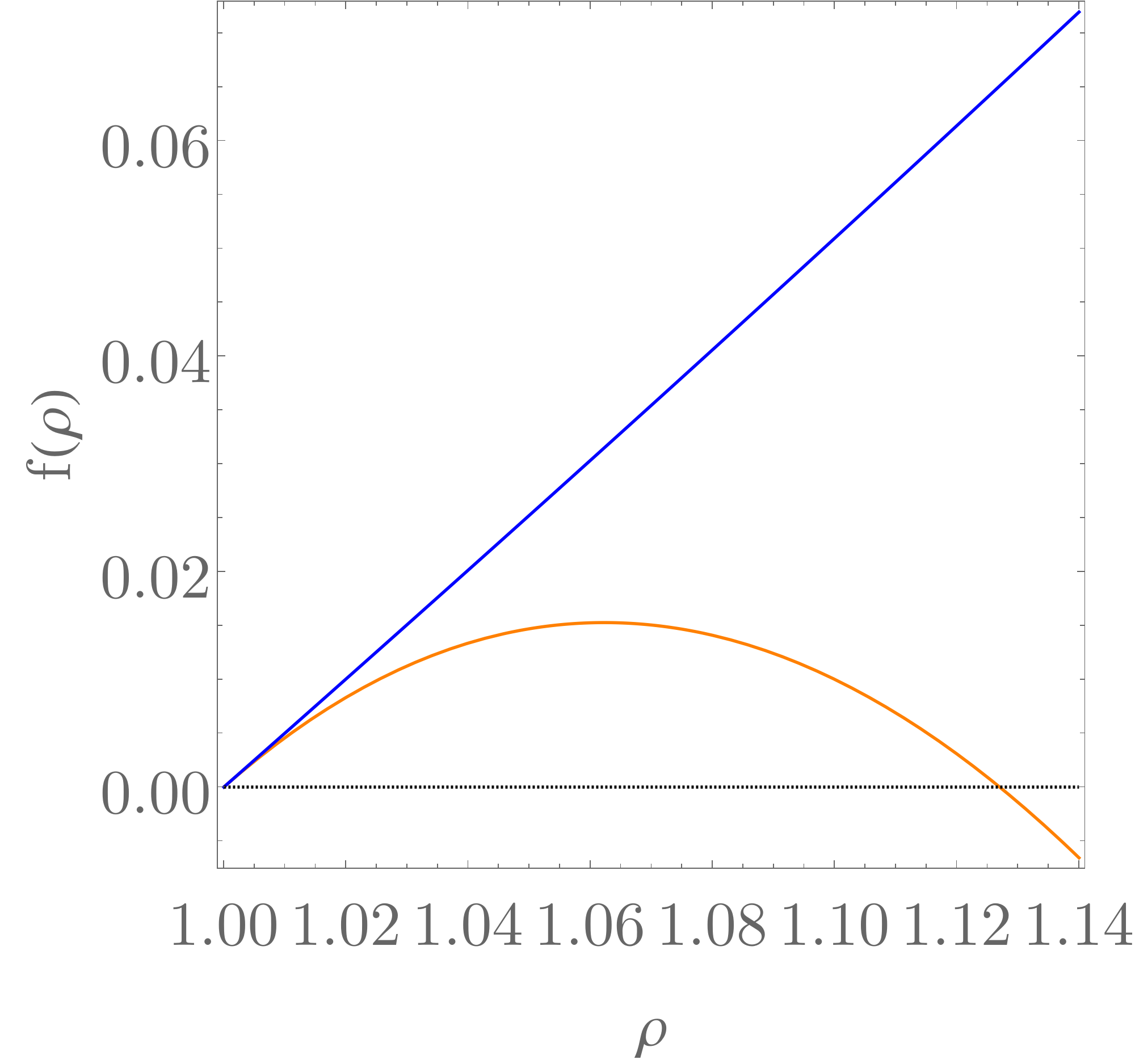}
\caption{Parameter space in the case of NUT. In the leftmost plot, the gray region corresponds to solutions with two roots to the equation $f(\rho) = 0$, with $\rho\equiv r/n$. The black line corresponds to solutions with only one root and the white region corresponds to solutions with $f(r) > 0$ for $r > n$. The orange curve is a generic representative of the gray region ($\ell^2 =n^2$, $b=-(5/2)n$), and the blue curve is a generic representative of the white region ($\ell^2 = 9n$, $b=-2n$). These two cases are shown in the second plot. The third plot is a zoom-in of the region around $f(\rho)=0$.}\label{fig1}
\end{figure}

Depending on the region in the parameter space, the metric function $f_n(r)$ could have a single root at $r=n$ and, in other cases, a second one at $r=r_+>n$. This produces a compact self-dual Taub-NUT metric. The different behavior of the solution depending on the region in the parameter space are given in Figure~\ref{fig1}.  

The Taub-bolt case, on the other hand, leads to an instanton with a horizon located at the largest root of the polynomial $f(r_h)=0$. Indeed, the condition~\eqref{bolt} implies that the following relation between the parameters must hold 
\begin{align}
    a = \frac{\left[r_h(\ell^2-n^2-b^2)-b\left(r_h^2+n^2\right)\right](r_h+b)^2}{\ell^2\left[n^2b+b^2r_h+3r_h^2(b+r_h)\right]}\equiv a_{\rm bolt}\,.
\end{align}
This condition, in turn, implies that metric function $f(r)$ and the scalar $\Lambda(r)$ are 
\begin{subequations}
    \begin{align}\notag
    f_b(r) &= -\frac{r(r+b)^2}{b\left(r^2-n^2 \right)} + \frac{(r+b)^3\left(r\,b+n^2 \right)}{b\ell^2\left(r^2-n^2 \right)} \\
    & - \frac{(r_h+b)^2\left[r\,b^2 + b\left(3r^2+n^2 \right)+3r^3 \right]\left[r_hb^2+b\left(r_h^2+n^2 \right) + r_h\left(n^2-l^2 \right) \right]}{b\ell^2\left(r^2-n^2 \right)\left[n^2b+b^2r_h+3r_h^2(b+r_h)\right] } \,,\\
    \Lambda_b(r) &= \frac{\left[r_h(\ell^2-n^2-b^2)-b\left(r_h^2+n^2\right)\right](r_h+b)^2}{\ell^2(r+b)^2\left[n^2b+b^2r_h+3r_h^2(b+r_h)\right]}\,, 
\end{align}
\end{subequations}
respectively. One can check that the condition $f_b(r_h)=0$ is satisfied. The absence of conical singularities at the bolt implies that $0\leq\tau\leq\beta_\tau$, where the period in Euclidean time is
\begin{align}
    \beta_\tau =  \frac{4\pi\ell^2\left[n^2b+b^2r_h+3r_h^2(r_h+b) \right]}{(r_h+b)\left[3r_h^2(r_h+b)+3r_h\left(\ell^2-n^2 \right)+b\left(\ell^2-3n^2 \right) \right]}\,.
\end{align}
The topology of the Taub-bolt instanton is different from that of Taub-NUT. The former connects a two-dimensional set of fixed points of the Killing vector $\partial_\tau$ with a locally Lorentz-flat asymptotic region $r\to\infty$ rather than a zero-dimensional set as in the case of NUT. The whole space has the topology of $\mathbb{CP}^2$ with a point removed and sent to infinity~\cite{Page:1978hdy}. Even though this space is not locally self-dual as in the Taub-NUT case, one can check that it is asymptotically locally self-dual in the sense of Eq.~\eqref{selfdualRcal}.

\subsection{Eguchi-Hanson-like solutions}  

There is another anti-self-dual gravitational instanton in the space of solutions of this theory. In the context of Einstein gravity, this solution was first found by Eguchi and Hanson in \cite{Eguchi:1978gw,Eguchi:1978xp}. For vanishing cosmological constant, it represents a Ricci-flat asymptotically locally Euclidean space with $\mathbb{R}^4/\mathbb{Z}_2$ topology; the solution with nonvanishing cosmological term was studied in Ref.~\cite{Pedersen_1985}. The hypersurfaces of constant radial coordinate have the topology of $\mathbb{S}^3/\mathbb{Z}_2=\mathbb{RP}^3$ and its isometry group is $U(1)\times SU(2)/\mathbb{Z}_2\simeq U(2)$. The Eguchi-Hanson (EH) metric was shown to be relevant in the resolution of orbifold singularities~\cite{Gibbons:1978tef,Gibbons:1979xm,Nitta:2020pzo} and it has been used to construct higher-dimensional solutions with nontrivial topology in Einstein gravity and beyond~\cite{Clarkson:2005qx,Clarkson:2006zk,Ishihara:2006pb,Chng:2006gh,Wong:2011aa,Wong:2012ne,Hendi:2012zg,Durgut:2022xzw}.

The generalized EH solution in the unconventional conformal supersymmetric theory studied here is found by choosing $h(r)=r^2/4$ in Eq.~\eqref{metricansatz}. Then, we find that the field equations are solved by $J^2(r)=h(r)$ and
\begin{subequations}\label{EHsolgeneral}
    \begin{align}
    f(r) &= 1-\frac{a^4}{r^4}-\frac{q^2}{r^2}+\frac{4pa^4\,r^2}{q^2}+pr^4 \,,\\
    \Lambda(r) &= -\frac{2\left(16pa^{12} - 4a^4q^4-q^8 \right)}{q^2\left(2a^4+q^2\,r^2\right)^2}\,. \,
\end{align}
\end{subequations}
where $a$, $q$, and $p$ are integration constants. This is a three-parameter family of solutions that can be regarded as a torsional extension of the metric found in Ref.~\cite{Corral:2021xsu}, in which the nontrivial components of the torsion for this solution are 
\begin{align}
    T^a &= - \frac{2q^2r\,\sqrt{f}}{2a^4+q^2r^2}\,e^a\wedge e^4\,.
\end{align}
Note that the torsion vanishes as $q\to0$ with $a\neq0$. In this sense, $q$ can be regarded as a torsional hair whose backreaction resembles that of the Maxwell field, as it happens in Poincaré gauge theories~\cite{Cembranos:2016gdt,Cembranos:2017pcs,Obukhov:2020hlp}. Now the components of the Lorentz curvature $2$-form are 
\begin{subequations}
    \begin{align}
    R^{12} &= \mathscr{W}(r)\,e^1\wedge e^2 - 2\mathscr{X}(r)\,e^3\wedge e^4\,, & R^{13} &= \mathscr{Y}(r)\,e^1\wedge e^3 - \mathscr{X}(r)\,e^2\wedge e^4\,, \\
    R^{14} &= \mathscr{Y}(r)\,e^1\wedge e^4 + \mathscr{X}(r)\,e^2\wedge e^3\,, & R^{23} &= \mathscr{Y}(r)\,e^2\wedge e^3 + \mathscr{X}(r)\,e^1\wedge e^4\,, \\
    R^{24} &= \mathscr{Y}(r)\,e^2\wedge e^4 - \mathscr{X}(r)\,e^1\wedge e^3\,, & R^{34} &= \mathscr{Z}(r)\,e^3\wedge e^4 - 2\mathscr{X}(r)\,e^1\wedge e^2\,,
\end{align}
\end{subequations}
where
\begin{subequations}
    \begin{align}
    \mathscr{W}(r) &= \frac{8fa^4q^2}{\left(2a^4+q^2r^2 \right)^2} - \frac{4\left(f-1 \right)}{r^2}\,, \\
    \mathscr{X}(r) &= \frac{f'}{2r} \,,\\
    \mathscr{Y}(r) &= \frac{16fa^4q^2r-f'\left(4a^8-q^4r^4 \right) }{2r\left(2a^4+q^2r^2 \right)^2}\,, \\
    \mathscr{Z}(r) &= - \frac{\left[f'r\left(2a^4+q^2r^2 \right)^2 \right]' + 2f'\left(2a^4-q^2r^2 \right)^2-8f'q^4r^4-16fa^4q^2r }{2r\left(2a^4+q^2r^2 \right)^2}\,.
\end{align}
\end{subequations}
If $p\neq0$, this solution has a curvature singularity for $r\to\infty$, while its action and topological invariants diverge. This can be seen from the asymptotic expansion of the torsion $2$-form for large $r$,
\begin{align}
    T^a &= - \left[2\sqrt{p}\;r + \frac{q^4-4a^8p}{q^4\sqrt{p}\;r^3} + \mathcal{O}(r^{-5}) \right]e^a\wedge e^4\,.
\end{align}
On the other hand, the fall-off of the Lorentz curvature $2$-form in this non-coordinate basis as $r\to\infty$ is characterized by 
\begin{subequations}\label{asympcurvEH}
\begin{align}
    \mathscr{W}(r) &= -4pr^2 - \frac{8pa^4}{q^2} + \frac{4\left(2a^4q^4+q^8-8a^{12}p \right)}{q^6 r^4} + \mathcal{O}(r^{-6})\,, \\
    \mathscr{X}(r) &= 2pr^2 + \frac{4pa^4}{q^2} + \frac{q^2}{r^4} + \mathcal{O}(r^{-6})\,, \\
    \mathscr{Y}(r) &= 2pr^2 + \frac{4pa^4}{q^2} + \frac{8a^4q^4+q^8-32a^{12}p}{q^6r^4} + \mathcal{O}(r^{-6})\,, \\
    \mathscr{Z}(r) &= -4pr^2 - \frac{8pa^4}{q^2} + \frac{4\left(2a^4q^4+q^8-8a^{12}p \right)}{q^6 r^4} + \mathcal{O}(r^{-6})\,.
\end{align}    
\end{subequations}
Thus, if $p\neq0$, this solution cannot be regarded as a gravitational instanton, whereas if $p=0$ the solution becomes
\begin{subequations}\label{solEHsd}
\begin{align}
    f(r) &= 1-\frac{a^4}{r^4}-\frac{q^2}{r^2} \,,\\
    \Lambda(r) &= \frac{2q^2\left(4a^4+q^4 \right)}{\left(2a^4+q^2\,r^2\right)^2}\,, \,
\end{align}    
\end{subequations}
and it is anti-self dual in the sense of Eq.~\eqref{selfdualRcal}. This solution has been studied in the context of Einstein gravity in Refs.~\cite{Xiao_2004,Chen:2020org,Yang:2023snt}. Indeed, the asymptotic behavior of the torsion in this case is given by
\begin{align}
    T^a &= -\left[\frac{2}{r} - \frac{q^4+4a^4}{q^2r^3} - \frac{\left(q^4+4a^4 \right)\left(q^4-8a^4\right)}{4q^4r^5} + \mathcal{O}(r^{-7}) \right]e^a\wedge e^4\,,
\end{align}
while the asymptotic expansion for the Lorentz curvature $2$-form can be read explicitly from Eq.~\eqref{asympcurvEH} by taking $p=0$. The solution of Eq.~\eqref{solEHsd} has a bolt at $r=r_h$ defined by the largest real positive root of the polynomial $f(r_h)=0$, that is,
\begin{align}
    r_h^2 = \frac{1}{2}\left(q^2+\sqrt{q^4+4a^2}\right)\,.
\end{align}
This bolt exists for all $a,q\in\mathbb{R}$ as the discriminant is always positive definite. The absence of conical singularities at the bolt, in turn, demands that the Euclidean-time coordinate has to be identified according to $0\leq\tau\leq\beta_\tau$, where
\begin{align}
   \beta_\tau = \frac{8\pi}{r_h\,f'|_{r=r_h}} = \frac{4\pi r_h^4}{2a^4 + q^2r_h^2} \,.
\end{align}
In the vanishing torsion limit, i.e. $q\to0$, the bolt is located at $r_h=a$ and the period of the Euclidean time is $2\pi$, giving a topology of $\mathbb{S}^3/\mathbb{Z}_2$ as in the original Eguchi-Hanson metric~\cite{Eguchi:1978gw,Eguchi:1978xp}.

Additionally, if $p\to0$ and $q\to0$ while keeping $p/q^2\equiv-\Lambda_0/(24a^4)$ fixed, the solution presented in Eq.~\eqref{EHsolgeneral} is continuously connected to the Eguchi-Hanson metric of general relativity with cosmological constant found in Ref.~\cite{Pedersen_1985}, that is,
\begin{align}\label{EHGR}
    f(r) = 1 - \frac{a^4}{r^4} - \frac{\Lambda_0\,r^2}{6}\;\;\;\;\; \mbox{and} \;\;\;\;\; \Lambda(r) = \Lambda_0\,,
\end{align}
where $\Lambda_0$ is an integration constant. Indeed, one can check using Eq.~\eqref{eqforLambda} that the torsion vanishes in this case. If $\Lambda_0\neq0$, the hypersurfaces of constant radius have the topology of $\mathbb{S}^3/\mathbb{Z}_k$ with $k$ an integer if $0\leq\tau\leq4\pi/k$ and
\begin{align}
r_h = \sqrt{-\frac{2(k-2)}{\Lambda_0}}\,.
\end{align}
If $\Lambda_0>0$, the existence of a bolt demands that $k<2$. In such a case, the radial coordinate is bounded since it appears a cosmological horizon. On the contrary, if $\Lambda_0<0$, then $k>2$, and the radial coordinate is unbounded from above. The case $\Lambda_0=0$ needs to be treated separately. In that case, the metric~\eqref{EHGR} becomes the standard EH metric of Ref.~\cite{Eguchi:1978gw,Eguchi:1978xp} with $r_h=a$ and $\mathbb{S}^3/\mathbb{Z}_2$ topology.

\section{Global properties}\label{sec:global} 

In this section, we obtain the global properties of the solutions such as their Euclidean on-shell actions and topological invariants. The first is an important quantity that allows us to obtain the partition function of the system to first order in the saddle-point approximation through the relation $\ln\mathcal{Z}\approx-I_E$. Here, we only consider (anti-)self-dual instantons in the sense of Eq.~\eqref{selfdualRcal} for the sake of simplicity, as the value of their Euclidean on-shell action is proportional to the Chern-Pontryagin index as we showed in Eq.~\eqref{IequalP1}. In the case of (anti-)self-dual Taub-NUT/AdS solution in Eq.~\eqref{solnut}, its action is
\begin{align}\label{Ionshell}
    I_{\rm on-shell} = 4\pi^2 P_1[\mathcal{M}] =  \frac{8\pi^2\left(\ell^2 - 4n^2 \right)\left(b^2+6nb^2+17n^2\right)}{\ell^4(b+3n)^2}\,.
\end{align}
Note that if $\ell=2n$ the Chern-Pontryagin index vanishes for the self-dual solution in Eq.~\eqref{solnut}. This is related to the fact that at this point in parameter space the solution becomes global (Euclidean) AdS with a nontrivial slicing \cite{Johnson:2014xza,Johnson:2014pwa}, which can be rendered as the ground state of the system. On the other hand, the Euclidean on-shell action of the torsional anti-self-dual EH instanton \eqref{solEHsd} is 
\begin{align}\label{actionTN}
    I_{\rm on-shell} = -4\pi^2 P_1[\mathcal{M}] = \frac{4\pi^2\left(q^4-4r_h^2q^2 +6r_h^4\right)}{r_h^2\left(q^2-2r_h^2\right)} \,.
\end{align}
In the vanishing torsion limit, i.e. $q\to0$, the action is equal to $12\pi^2$ giving a Chern-Pontryagin index $P_1[\mathcal{M}]=-3$, a well-known result of the original EH space~\cite{Eguchi:1978gw,Eguchi:1978xp}.

The Euler characteristic of these configurations can be computed as~\cite{Eguchi:1980jx}
\begin{align}\label{Euler}
    \chi(\mathcal{M}) &= -\frac{1}{32\pi^2}\left[\int_{\mathcal{M}}\epsilon_{abcd}R^{ab}\wedge R^{cd} - 2\int_{\partial\mathcal{M}}\epsilon_{abcd}\,\theta^{ab}\wedge\left(R^{cd} - \frac{2}{3}\,\theta^{a}_{\ c}\wedge\theta^{cd}\right)\right]\,, 
\end{align}
where $\theta^{ab}=\omega^{ab}-\omega_0^{ab}$ is the second fundamental form possessing only normal components to the boundary $\partial\mathcal{M}$ and $\omega_0^{ab}$ denotes the Lorentz connection of the product space having only tangential components at the boundary. To compute the second fundamental form, we evaluate the torsional components at $r=r_0$ with $r_0$ constant, such that the full Lorentz connection $\omega_0^{ab}$ has tangential components at the boundary. In the case of the novel torsional self-dual Taub-NUT/AdS of Eq.~\eqref{solnut}, the nontrivial components of $\theta^{ab}$ are given by
\begin{subequations}
    \begin{align}
\theta^{14} &= \frac{(br_0+n^2)\Phi(r_0)}{\ell\left(r_0^2-n^2 \right)(b+r_0)}\, e^1 \,, \\
\theta^{24} &= \frac{(br_0+n^2)\Phi(r_0)}{\ell\left(r_0^2-n^2 \right)(b+r_0)}\, e^2\,, \\
\theta^{34} &= -\left[\frac{(br_0+n^2)\Phi(r_0)}{\ell\left(r_0^2-n^2 \right)(b+r_0)}-\frac{\left(\ell^2-4n^2 \right)(r_0-n)(b+r_0)}{\ell(r_0+n)^2(3n+b)\Phi(r_0)}\right]e^3\,,
\end{align}
\end{subequations}
where we have defined
\begin{align}
    \Phi(r_0) = \sqrt{\frac{(r_0-n)\left[b(r_0^2+\ell^2) + 2\ell^2r_0 +n(3r_0^2+\ell^2+2br_0) -n^2(2r_0+3b)-n^3\right]}{(r_0+n)(3n+b)}}\,.
\end{align}
Using this result, direct evaluation of the Euler characteristic for the self-dual Taub-NUT/AdS instanton yields
\begin{align}
    \chi(\mathcal{M}) = 1\,.
\end{align}
A similar procedure carried out for the Taub-bolt instanton gives $\chi(\mathcal{M})=2$. Then, the Euler characteristics of the Taub-NUT and Taub-bolt spaces differ by one unit, which reflects their different number of boundaries. Moreover, these values match those of the Riemannian case, showing that the affine properties of these spaces do not change the topological invariants of the Euler class. 

In the case of the anti-self-dual EH solution, we follow the same strategy: we consider a product space of constant radial coordinate, i.e. $r=r_0$, such that the Lorentz connection has only tangential components at the boundary. Then, the nontrivial components of the second-fundamental form are
\begin{subequations}
\begin{align}
    \theta^{14} &= \frac{\left(2a^4-q^2r_0^2\right)\sqrt{r_0^4-q^2r_0^2-a^4}}{r_0^3\left(2a^4+q^2r_0^2\right)}\,e^1\,, \\
    \theta^{24} &= \frac{\left(2a^4-q^2r_0^2\right)\sqrt{r_0^4-q^2r_0^2-a^4}}{r_0^3\left(2a^4+q^2r_0^2\right)}\,e^2\,, \\
    \theta^{34} &= \frac{2a^8+3a^4q^2r_0^2+2a^4r_0^4+2q^4r_0^4-q^2r_0^6}{r_0^3\left(2a^4+q^2r_0^2\right)\sqrt{r_0^4-q^2r_0^2-a^4}}\,e^3\,.
\end{align}    
\end{subequations}
Then, definition \eqref{Euler} yields the Euler characterstic of the torsional extension of the EH geometry as
\begin{align}
    \chi(\mathcal{M}) = 2\,.
\end{align}

Computing the index theorem for the Dirac operator is more involved. For non-compact manifolds, one needs to introduce the codimension-1 boundary integral of the Chern-Simons form that can be evaluated directly from $\int_{\partial\mathcal{M}}\theta^{a}_{\ b}\wedge R^{b}_{\ a}$. Additionally, one needs to include the Atiyah-Patodi-Singer $\eta$-invariant that measures the difference between the positive and negative eigenvalues of the tangential components of the Dirac operator~\cite{APS}. In absence of torsion, one can use the Hitchin formula~\cite{HITCHIN19741} to compute the $\eta$-invariant directly. Nevertheless, in the presence of torsion, this formula does not apply and one needs to evaluate this invariant from its definition by using the Riemann-zeta function regularization procedure~\cite{APS}. Moreover, the Dirac index receives corrections coming from the Nieh-Yan invariant~\cite{Hojman:1980kv,Nelson:1980ph,Nieh:1981ww}, since it has been shown that this term contributes nontrivially to the axial anomaly~\cite{Chandia:1997hu}. This is certainly an interesting aspect that goes beyond the scope of this work and we plan to revisit it in the future.  

\section{Summary}\label{sec:summary}

In this work we have considered the definition of (anti-)self duality brought in by unconventional supersymmetry. In the bosonic sector of the theory, this (anti-)self duality corresponds to conditions on the bosonic curvature components that saturate a BPS bound and, if the action is augmented by a topological term of the Pontryagin class, the moduli space of vacuum solutions of the theory is enlarged to include all the (anti-)self-dual configurations as ground states. 

In the bosonic sector of the theory, we obtained explicit solutions of conformal unconventional supersymmetry that are (anti-)self dual. In particular, we focused on gravitational instantons constructed out of the $U(1)$-fibration of Einstein-K\"ahler manifolds that generalize the Taub-NUT/Bolt-AdS and Eguchi-Hanson spaces of general relativity including torsion. In both cases, the torsion is sourced by the compensating fields associated to the AdS boosts and special conformal transformations, but these fields are auxiliary and do not represent propagating deegrees of freedom. These solutions are shown to be continuously connected to their Einsteinian counterparts in the vanishing torsion limit.

Additionally, the global properties of these solutions such as the action and their topological invariants have been computed. In the bosonic sector, the action turns out to be proportional to the Chern-Pontryagin index of $SO(4)$ that is explicitly computed; this is usually the case for instantons in Yang-Mills theory. Although the geometry of these solutions is different from that of their Riemannian counterparts as they include nontrivial torsion, the value of the Euler characteristic is insensitive to torsion, showing that the different affine properties of these spaces do not modify this aspect of their topology.

Interesting questions remain open. For instance, the contribution of these (anti-)self-dual configurations to the Dirac index is certainly of great interest. This could provide a connection between torsion and the axial anomaly as shown in \cite{Chandia:1997hu}. Moreover, since the solutions presented here are non-compact, one would expect a boundary contribution to the topological invariants of the Pontryagin class. Although bulk contributions of torsion to the axial anomaly have been explored, the presence of boundaries implies that an extended analysis is needed. Finally, since these configurations saturate a BPS bound, it can be expected that they preserve a fraction of the supersymmetry only. We plan to study the existence of Killing spinors in these spaces to address this aspect in the future.     

\begin{acknowledgments}

We thank to Fabrizio Canfora, Oscar Castillo-Felisola, Nelson Merino, Omar Valdivia, and Mauricio Valenzuela for insightful comments and discussions.  This work has been partially funded by Agencia Nacional de Investigaci\'{o}n y Desarrollo (ANID) through FONDECYT grants ~11200025, 1201208, 1210500, 1220862 and~1230112. 

\end{acknowledgments}

\appendix

\section{Fundamental representation of $SU(2,2|N)$}\label{sec:representationApp}

Throughout this work, we consider the following representation of $SU(2,2|N)$
\begin{align}
    (\JJ_a)^A_{\ B}&=\frac{1}{2}(\gamma_a)^\alpha_{\ \beta}\delta^A_{\ \alpha} \delta^\beta_{\ B}=\frac{1}{2}(\gamma_a)^A_{\ B}\,, &
    (\JJ_{ab})^A_{\ B}&=\frac{1}{4}[\gamma_a,\gamma_b]^A_{\ B}=(\Sigma_{ab})^A_{\ B}\,, \\
    (\KK_a)^A_{\ B}&=\frac{1}{2}(\tilde{\gamma}_a)^A_{\ B}\,, &
    (\DD)^A_{\ B}&=\frac{1}{2}(\gamma_5)^A_{\ B}\,,\\
    (\TT_I)^A_{\ B}&=\frac{i}{2}(\lambda_I^{t})_{\ B}^{A} \,,&
    (\QQ^\alpha_i)^A_{\ B}&=\delta^A_{i} \delta^\alpha_B\,,\\
    (\QQb_\alpha^i)^A_{\ B}&= \delta^A_\alpha \delta^i_B\,,&
    \ZZ^A_{\ B} &= \left(i\delta^A_\alpha\delta^\alpha_B+\frac{4i}{N}\delta^A_i\delta^i_B\right)\,,
\end{align}
where $\gamma_5=i\gamma^0 \gamma^1 \gamma^2 \gamma^3$, satisfying $(\gamma_5)^2=\mathds{1}$, and we have defined $\tilde{\gamma}^a=-\gamma_5\gamma^a$.
The $\gamma$-matrices are in a $4\times 4$ spinor-representation ($\alpha, \beta,\cdots$ run from 1 to 4). The indices of the tangent space are $a,b=0,1,2,3$ while indices in the adjoint representation of $\mathfrak{su}(N)$ take values $I,J=1,2,\ldots,N^{2}-1$, and in the fundamental representation they take the values $i,j=1,2,\ldots,N$. The $\gamma$-matrices are endomorphisms and they act on spinors as
\begin{equation}
 \psi^\alpha \stackrel{\gamma_a}{\longrightarrow} (\gamma_a)^\alpha_{\ \beta} \psi^\beta\,.
\end{equation}

In a similar way the $\lambda$-matrices are also endomorphisms and they act on spinors as
\begin{equation}
 \psi^\alpha_i \stackrel{\lambda_I}{\longrightarrow} (\lambda_I)_i^{\ j}\psi^\alpha_j \,,
\end{equation}
and they satisfy the algebra $[\lambda_I,\lambda_J]=f_{IJK} \lambda^K$, where indexes are raised/lowered with an Euclidean metric $\delta_{IJ}$. Indices of the adjoint representation of $SU(2,2|N)$ are denoted by $A,B=1,\cdots,N+4$, so we have a $N+4\times N+4$ representation. We find convenient to split them as $A=(\alpha,i)$. Then, all the possible products that mix spaces like $p^i_{\ A}  q^A_{\ \alpha}$ are trivial. Thus, the following relations are understood
\begin{align}
(\gamma_a)^A_{\ B}=\delta^A_\alpha (\gamma_a)^\alpha_{\ \beta} \delta^\beta_B \;\;\;\;\; \mbox{and} \;\;\;\;\; C_{\alpha A}=C_{\alpha\beta} \delta^\beta_A \,.
\end{align}

The generators $\JJ_a$ and $\JJ_{ab}$ form a AdS$_4$ algebra and, alongside $\DD$ and $\KK_a$, they form the conformal algebra, that is,
\begin{align}
[\JJ_a,\JJ_b]&=\JJ_{ab}\,, &
[\JJ_a,\JJ_{bc}]&=\eta_{ab}\JJ_c-\eta_{ac}\JJ_b\,,\\
[\JJ_{ab},\JJ_{cd}]&=-(\eta_{ac}\JJ_{bd}-\eta_{ad}\JJ_{bc}-\eta_{bc}\JJ_{ad}+\eta_{bd}\JJ_{ac})\,, &
[\KK_a,\KK_b]&=-\JJ_{ab}\,, \\
[\JJ_a,\KK_{b}]&=\eta_{ab}\DD\,,&
[\KK_a,\JJ_{bc}]&=\eta_{ab}\KK_c-\eta_{ac}\KK_b\,,\\
[\DD,\KK_{a}]&=-\JJ_a\,,&
[\DD,\JJ_{a}]&=-\KK_a\,.
\end{align}

Including $\QQ^\alpha_i$ and $\QQb^i_\alpha$, the commutators close in a $\mathfrak{su}(2,2|N)$ superalgebra given by
\begin{align}
[\JJ_a,\QQb_\alpha^i]&=\frac{1}{2}\QQb_\beta^i(\gamma_a)^\beta_{\ \alpha}\,, & [\JJ_a,\QQ^\alpha_i]&=-\frac{1}{2}(\gamma_a)^\alpha_{\ \beta}\QQ^\beta_i\,,\\
[\JJ_{ab},\QQb_\alpha^i]&=\QQb_\beta^i(\Sigma_{ab})^\beta_{\ \alpha}\,, & [\JJ_{ab},\QQ^\alpha_i]&=-(\Sigma_{ab})^\alpha_{\ \beta}\QQ^\beta_i\,,\\
[\KK_a,\QQb_\alpha^i]&=\frac{1}{2}\QQb_\beta^i(\tilde{\gamma}_a)^\beta_{\ \alpha}\,, & [\KK_a,\QQ^\alpha_i]&=-\frac{1}{2}(\tilde{\gamma}_a)^\alpha_{\ \beta}\QQ^\beta_i\,,\\
[\DD,\QQb_\alpha^i]&=\frac{1}{2}\QQb_\beta^i(\gamma_5)^\beta_{\ \alpha}\,, & [\DD,\QQ^\alpha_i]&=-\frac{1}{2}(\gamma_5)^\alpha_{\ \beta}\QQ^\beta_i\,,\\
[\TT_I,\QQb_\alpha^i]&=-\frac{i}{2}\QQb_\alpha^j (\lambda_I)^{\ i}_j\,, & [\TT_I,\QQ^\alpha_i]&=\frac{i}{2}(\lambda_I)^{\ j}_i\QQ^\alpha_j\,,\\
[\ZZ,\QQb_\alpha^i]&=-i(4/N-1)\QQb_\alpha^i\,, & [\ZZ,\QQ^\alpha_i]&=i(4/N-1)\QQ^\alpha_i\,,
\end{align}
and 
\begin{align}
    \{\QQ^\alpha_i,\QQb_\beta^j\}&=\left(\frac{1}{2}(\gamma^a)^\alpha_{\ \beta} \JJ_a-\frac{1}{2}(\Sigma^{ab})^\alpha_{\ \beta} \JJ_{ab}-\frac{1}{2}(\tilde{\gamma}^a)^\alpha_{\ \beta} \KK_a+\frac{1}{2}(\gamma_5)^\alpha_{\ \beta} \DD\right)\delta^j_i \nonumber\\
&\qquad \qquad \qquad +\delta^\alpha_\beta\left(-i(\lambda_I)_i^{\ j}\TT_I-\frac{i}{4}\delta_i^{j} \ZZ\right)\,. 
\end{align}

The graduation operator is given by
\begin{equation}
\mathcal{G}^A_{\ B} = \delta^A_{\alpha}\delta^\alpha_{\ B}-\delta^{A}_{i}\delta^i_{\ B}\,,
\end{equation}
and it classifies generators in bosonic $B=\{\JJ_a,\JJ_{ab},\KK_a,\DD,\TT_I,\ZZ\}$ or fermionic $F=\{Q^\alpha_i,\QQb^i_\alpha\}$ representations by $[B,\mathcal{G}]=0=\{F,\mathcal{G}\}$, and it squares to one, i.e., $\mathcal{G}^2=1$. With the graduation operator we can define an invariant supertrace
\begin{equation}
\langle G\rangle  \equiv \Tr(\mathcal{G} G)=0\,.
\end{equation}
The supertrace have the following properties
\begin{equation}
 \langle B_1 B_2\rangle =\langle B_2 B_1\rangle \,, \quad \langle B F\rangle =\langle F B \rangle \,, \quad \langle F_1 F_2\rangle =-\langle F_2 F_1\rangle \,.
\end{equation}

All the generators $G$ in the representation are s-traceless
\begin{equation}
\langle G\rangle=0\,, \;\;\;\;\; \mbox{where} \;\;\;\;\; G=\{\JJ_a,\JJ_{ab},\KK_a,\DD,\TT_I,\ZZ,Q^\alpha_i,\QQb^i_\alpha\}\,.
\end{equation}
The quadratic combinations that give nontrivial traces are
\begin{align}
\langle \JJ_a \JJ_b \rangle&=\eta_{ab}\,, & \langle \JJ_{ab} \JJ_{cd} \rangle&=-(\eta_{ac}\eta_{bd}-\eta_{bc}\eta_{ad})\,,\label{trace1}\\
 \langle \KK_a \KK_b \rangle&=-\eta_{ab}\,, & \langle \DD^2\rangle &= +1\,,\\
\langle \TT_I \TT_J \rangle&=\frac{1}{2}\delta_{IJ}\,, & \langle \ZZ^2\rangle&=4(4/N-1)\,,\\
\langle \QQ^\alpha_i \QQb^j_\beta\rangle&=-\delta^\alpha_\beta \delta^j_i=-\langle \QQb^j_\beta \QQ^\alpha_i\rangle\,.&\label{trace-1}
\end{align}

The $S$ operator is fundamental in getting the usual expressions for the kinetic terms. In  this case, the nontrivial traces are
\begin{align}
 \langle S \DD \rangle &=2i\,, &
 \langle S \JJ_{ab}\JJ_{cd}\rangle&=-\epsilon_{abcd}=\langle \JJ_{ab}S\JJ_{cd}\rangle\,,\\
 \langle  \JJ_{a}S\KK_{b}\rangle&=-i  \eta_{ab}=-\langle \KK_{a}S\JJ_{b}\rangle\,,&
 \langle \ZZ S \DD \rangle&=-2=\langle \DD S \ZZ \rangle\,.
\end{align}

\bibliography{paper_selfdual.bib}

\begin{thebibliography}{77}%
\makeatletter
\providecommand \@ifxundefined [1]{%
 \@ifx{#1\undefined}
}%
\providecommand \@ifnum [1]{%
 \ifnum #1\expandafter \@firstoftwo
 \else \expandafter \@secondoftwo
 \fi
}%
\providecommand \@ifx [1]{%
 \ifx #1\expandafter \@firstoftwo
 \else \expandafter \@secondoftwo
 \fi
}%
\providecommand \natexlab [1]{#1}%
\providecommand \enquote  [1]{``#1''}%
\providecommand \bibnamefont  [1]{#1}%
\providecommand \bibfnamefont [1]{#1}%
\providecommand \citenamefont [1]{#1}%
\providecommand \href@noop [0]{\@secondoftwo}%
\providecommand \href [0]{\begingroup \@sanitize@url \@href}%
\providecommand \@href[1]{\@@startlink{#1}\@@href}%
\providecommand \@@href[1]{\endgroup#1\@@endlink}%
\providecommand \@sanitize@url [0]{\catcode `\\12\catcode `\$12\catcode
  `\&12\catcode `\#12\catcode `\^12\catcode `\_12\catcode `\%12\relax}%
\providecommand \@@startlink[1]{}%
\providecommand \@@endlink[0]{}%
\providecommand \url  [0]{\begingroup\@sanitize@url \@url }%
\providecommand \@url [1]{\endgroup\@href {#1}{\urlprefix }}%
\providecommand \urlprefix  [0]{URL }%
\providecommand \Eprint [0]{\href }%
\providecommand \doibase [0]{http://dx.doi.org/}%
\providecommand \selectlanguage [0]{\@gobble}%
\providecommand \bibinfo  [0]{\@secondoftwo}%
\providecommand \bibfield  [0]{\@secondoftwo}%
\providecommand \translation [1]{[#1]}%
\providecommand \BibitemOpen [0]{}%
\providecommand \bibitemStop [0]{}%
\providecommand \bibitemNoStop [0]{.\EOS\space}%
\providecommand \EOS [0]{\spacefactor3000\relax}%
\providecommand \BibitemShut  [1]{\csname bibitem#1\endcsname}%
\let\auto@bib@innerbib\@empty
\bibitem [{\citenamefont {Alvarez}\ \emph {et~al.}(2012)\citenamefont
  {Alvarez}, \citenamefont {Valenzuela},\ and\ \citenamefont
  {Zanelli}}]{Alvarez:2011gd}%
  \BibitemOpen
  \bibfield  {author} {\bibinfo {author} {\bibfnamefont {P.~D.}\ \bibnamefont
  {Alvarez}}, \bibinfo {author} {\bibfnamefont {M.}~\bibnamefont {Valenzuela}},
  \ and\ \bibinfo {author} {\bibfnamefont {J.}~\bibnamefont {Zanelli}},\ }\href
  {\doibase 10.1007/JHEP04(2012)058} {\bibfield  {journal} {\bibinfo  {journal}
  {JHEP}\ }\textbf {\bibinfo {volume} {04}},\ \bibinfo {pages} {058} (\bibinfo
  {year} {2012})}\BibitemShut {NoStop}%
\bibitem [{\citenamefont {Alvarez}\ \emph
  {et~al.}(2014{\natexlab{a}})\citenamefont {Alvarez}, \citenamefont {Pais},\
  and\ \citenamefont {Zanelli}}]{Alvarez:2013tga}%
  \BibitemOpen
  \bibfield  {author} {\bibinfo {author} {\bibfnamefont {P.~D.}\ \bibnamefont
  {Alvarez}}, \bibinfo {author} {\bibfnamefont {P.}~\bibnamefont {Pais}}, \
  and\ \bibinfo {author} {\bibfnamefont {J.}~\bibnamefont {Zanelli}},\ }\href
  {\doibase 10.1016/j.physletb.2014.06.031} {\bibfield  {journal} {\bibinfo
  {journal} {Phys. Lett. B}\ }\textbf {\bibinfo {volume} {735}},\ \bibinfo
  {pages} {314} (\bibinfo {year} {2014}{\natexlab{a}})}\BibitemShut {NoStop}%
\bibitem [{\citenamefont {Alvarez}\ \emph {et~al.}(2022)\citenamefont
  {Alvarez}, \citenamefont {Chavez},\ and\ \citenamefont
  {Zanelli}}]{Alvarez:2021zsw}%
  \BibitemOpen
  \bibfield  {author} {\bibinfo {author} {\bibfnamefont {P.~D.}\ \bibnamefont
  {Alvarez}}, \bibinfo {author} {\bibfnamefont {R.~A.}\ \bibnamefont {Chavez}},
  \ and\ \bibinfo {author} {\bibfnamefont {J.}~\bibnamefont {Zanelli}},\ }\href
  {\doibase 10.1007/JHEP02(2022)111} {\bibfield  {journal} {\bibinfo  {journal}
  {JHEP}\ }\textbf {\bibinfo {volume} {02}},\ \bibinfo {pages} {111} (\bibinfo
  {year} {2022})}\BibitemShut {NoStop}%
\bibitem [{\citenamefont {Alvarez}\ \emph {et~al.}(2020)\citenamefont
  {Alvarez}, \citenamefont {Valenzuela},\ and\ \citenamefont
  {Zanelli}}]{Alvarez:2020qmy}%
  \BibitemOpen
  \bibfield  {author} {\bibinfo {author} {\bibfnamefont {P.~D.}\ \bibnamefont
  {Alvarez}}, \bibinfo {author} {\bibfnamefont {M.}~\bibnamefont {Valenzuela}},
  \ and\ \bibinfo {author} {\bibfnamefont {J.}~\bibnamefont {Zanelli}},\ }\href
  {\doibase 10.1007/JHEP07(2020)205} {\bibfield  {journal} {\bibinfo  {journal}
  {JHEP}\ }\textbf {\bibinfo {volume} {07}},\ \bibinfo {pages} {205} (\bibinfo
  {year} {2020})}\BibitemShut {NoStop}%
\bibitem [{\citenamefont {Alvarez}\ \emph {et~al.}(2021)\citenamefont
  {Alvarez}, \citenamefont {Delage}, \citenamefont {Valenzuela},\ and\
  \citenamefont {Zanelli}}]{Alvarez:2021qbu}%
  \BibitemOpen
  \bibfield  {author} {\bibinfo {author} {\bibfnamefont {P.~D.}\ \bibnamefont
  {Alvarez}}, \bibinfo {author} {\bibfnamefont {L.}~\bibnamefont {Delage}},
  \bibinfo {author} {\bibfnamefont {M.}~\bibnamefont {Valenzuela}}, \ and\
  \bibinfo {author} {\bibfnamefont {J.}~\bibnamefont {Zanelli}},\ }\href
  {\doibase 10.1007/JHEP07(2021)176} {\bibfield  {journal} {\bibinfo  {journal}
  {JHEP}\ }\textbf {\bibinfo {volume} {07}},\ \bibinfo {pages} {176} (\bibinfo
  {year} {2021})}\BibitemShut {NoStop}%
\bibitem [{\citenamefont {Alvarez}\ \emph
  {et~al.}(2023{\natexlab{a}})\citenamefont {Alvarez}, \citenamefont {Chavez},\
  and\ \citenamefont {Zanelli}}]{Alvarez:2022eew}%
  \BibitemOpen
  \bibfield  {author} {\bibinfo {author} {\bibfnamefont {P.~D.}\ \bibnamefont
  {Alvarez}}, \bibinfo {author} {\bibfnamefont {R.~A.}\ \bibnamefont {Chavez}},
  \ and\ \bibinfo {author} {\bibfnamefont {J.}~\bibnamefont {Zanelli}},\ }\href
  {\doibase 10.1007/JHEP02(2023)050} {\bibfield  {journal} {\bibinfo  {journal}
  {JHEP}\ }\textbf {\bibinfo {volume} {02}},\ \bibinfo {pages} {050} (\bibinfo
  {year} {2023}{\natexlab{a}})}\BibitemShut {NoStop}%
\bibitem [{\citenamefont {Belavin}\ \emph {et~al.}(1975)\citenamefont
  {Belavin}, \citenamefont {Polyakov}, \citenamefont {Schwartz},\ and\
  \citenamefont {Tyupkin}}]{Belavin:1975fg}%
  \BibitemOpen
  \bibfield  {author} {\bibinfo {author} {\bibfnamefont {A.~A.}\ \bibnamefont
  {Belavin}}, \bibinfo {author} {\bibfnamefont {A.~M.}\ \bibnamefont
  {Polyakov}}, \bibinfo {author} {\bibfnamefont {A.~S.}\ \bibnamefont
  {Schwartz}}, \ and\ \bibinfo {author} {\bibfnamefont {Y.~S.}\ \bibnamefont
  {Tyupkin}},\ }\href {\doibase 10.1016/0370-2693(75)90163-X} {\bibfield
  {journal} {\bibinfo  {journal} {Phys. Lett. B}\ }\textbf {\bibinfo {volume}
  {59}},\ \bibinfo {pages} {85} (\bibinfo {year} {1975})}\BibitemShut {NoStop}%
\bibitem [{\citenamefont {'t~Hooft}(1976{\natexlab{a}})}]{tHooft:1976rip}%
  \BibitemOpen
  \bibfield  {author} {\bibinfo {author} {\bibfnamefont {G.}~\bibnamefont
  {'t~Hooft}},\ }\href {\doibase 10.1103/PhysRevLett.37.8} {\bibfield
  {journal} {\bibinfo  {journal} {Phys. Rev. Lett.}\ }\textbf {\bibinfo
  {volume} {37}},\ \bibinfo {pages} {8} (\bibinfo {year}
  {1976}{\natexlab{a}})}\BibitemShut {NoStop}%
\bibitem [{\citenamefont {'t~Hooft}(1976{\natexlab{b}})}]{tHooft:1976snw}%
  \BibitemOpen
  \bibfield  {author} {\bibinfo {author} {\bibfnamefont {G.}~\bibnamefont
  {'t~Hooft}},\ }\href {\doibase 10.1103/PhysRevD.14.3432} {\bibfield
  {journal} {\bibinfo  {journal} {Phys. Rev. D}\ }\textbf {\bibinfo {volume}
  {14}},\ \bibinfo {pages} {3432} (\bibinfo {year} {1976}{\natexlab{b}})},\
  \bibinfo {note} {[Erratum: Phys. Rev. D 18, 2199 (1978)]}\BibitemShut
  {NoStop}%
\bibitem [{\citenamefont {Diakonov}\ and\ \citenamefont
  {Petrov}(1986)}]{Diakonov:1985eg}%
  \BibitemOpen
  \bibfield  {author} {\bibinfo {author} {\bibfnamefont {D.}~\bibnamefont
  {Diakonov}}\ and\ \bibinfo {author} {\bibfnamefont {V.~Y.}\ \bibnamefont
  {Petrov}},\ }\href {\doibase 10.1016/0550-3213(86)90011-8} {\bibfield
  {journal} {\bibinfo  {journal} {Nucl. Phys. B}\ }\textbf {\bibinfo {volume}
  {272}},\ \bibinfo {pages} {457} (\bibinfo {year} {1986})}\BibitemShut
  {NoStop}%
\bibitem [{\citenamefont {Eguchi}\ \emph {et~al.}(1980)\citenamefont {Eguchi},
  \citenamefont {Gilkey},\ and\ \citenamefont {Hanson}}]{Eguchi:1980jx}%
  \BibitemOpen
  \bibfield  {author} {\bibinfo {author} {\bibfnamefont {T.}~\bibnamefont
  {Eguchi}}, \bibinfo {author} {\bibfnamefont {P.~B.}\ \bibnamefont {Gilkey}},
  \ and\ \bibinfo {author} {\bibfnamefont {A.~J.}\ \bibnamefont {Hanson}},\
  }\href {\doibase 10.1016/0370-1573(80)90130-1} {\bibfield  {journal}
  {\bibinfo  {journal} {Phys. Rept.}\ }\textbf {\bibinfo {volume} {66}},\
  \bibinfo {pages} {213} (\bibinfo {year} {1980})}\BibitemShut {NoStop}%
\bibitem [{\citenamefont {Taub}(1951)}]{Taub:1950ez}%
  \BibitemOpen
  \bibfield  {author} {\bibinfo {author} {\bibfnamefont {A.~H.}\ \bibnamefont
  {Taub}},\ }\href {\doibase 10.2307/1969567} {\bibfield  {journal} {\bibinfo
  {journal} {Annals Math.}\ }\textbf {\bibinfo {volume} {53}},\ \bibinfo
  {pages} {472} (\bibinfo {year} {1951})}\BibitemShut {NoStop}%
\bibitem [{\citenamefont {Newman}\ \emph {et~al.}(1963)\citenamefont {Newman},
  \citenamefont {Tamburino},\ and\ \citenamefont {Unti}}]{Newman:1963yy}%
  \BibitemOpen
  \bibfield  {author} {\bibinfo {author} {\bibfnamefont {E.}~\bibnamefont
  {Newman}}, \bibinfo {author} {\bibfnamefont {L.}~\bibnamefont {Tamburino}}, \
  and\ \bibinfo {author} {\bibfnamefont {T.}~\bibnamefont {Unti}},\ }\href
  {\doibase 10.1063/1.1704018} {\bibfield  {journal} {\bibinfo  {journal} {J.
  Math. Phys.}\ }\textbf {\bibinfo {volume} {4}},\ \bibinfo {pages} {915}
  (\bibinfo {year} {1963})}\BibitemShut {NoStop}%
\bibitem [{\citenamefont {Eguchi}\ and\ \citenamefont
  {Hanson}(1978)}]{Eguchi:1978xp}%
  \BibitemOpen
  \bibfield  {author} {\bibinfo {author} {\bibfnamefont {T.}~\bibnamefont
  {Eguchi}}\ and\ \bibinfo {author} {\bibfnamefont {A.~J.}\ \bibnamefont
  {Hanson}},\ }\href {\doibase 10.1016/0370-2693(78)90566-X} {\bibfield
  {journal} {\bibinfo  {journal} {Phys. Lett. B}\ }\textbf {\bibinfo {volume}
  {74}},\ \bibinfo {pages} {249} (\bibinfo {year} {1978})}\BibitemShut
  {NoStop}%
\bibitem [{\citenamefont {Eguchi}\ and\ \citenamefont
  {Hanson}(1979)}]{Eguchi:1978gw}%
  \BibitemOpen
  \bibfield  {author} {\bibinfo {author} {\bibfnamefont {T.}~\bibnamefont
  {Eguchi}}\ and\ \bibinfo {author} {\bibfnamefont {A.~J.}\ \bibnamefont
  {Hanson}},\ }\href {\doibase 10.1016/0003-4916(79)90282-3} {\bibfield
  {journal} {\bibinfo  {journal} {Annals Phys.}\ }\textbf {\bibinfo {volume}
  {120}},\ \bibinfo {pages} {82} (\bibinfo {year} {1979})}\BibitemShut
  {NoStop}%
\bibitem [{\citenamefont {Gibbons}\ and\ \citenamefont
  {Hawking}(1978)}]{Gibbons:1978tef}%
  \BibitemOpen
  \bibfield  {author} {\bibinfo {author} {\bibfnamefont {G.~W.}\ \bibnamefont
  {Gibbons}}\ and\ \bibinfo {author} {\bibfnamefont {S.~W.}\ \bibnamefont
  {Hawking}},\ }\href {\doibase 10.1016/0370-2693(78)90478-1} {\bibfield
  {journal} {\bibinfo  {journal} {Phys. Lett. B}\ }\textbf {\bibinfo {volume}
  {78}},\ \bibinfo {pages} {430} (\bibinfo {year} {1978})}\BibitemShut
  {NoStop}%
\bibitem [{\citenamefont {Hitchin}(1979)}]{Hitchin:1979rts}%
  \BibitemOpen
  \bibfield  {author} {\bibinfo {author} {\bibfnamefont {N.~J.}\ \bibnamefont
  {Hitchin}},\ }\href {\doibase 10.1017/S0305004100055924} {\bibfield
  {journal} {\bibinfo  {journal} {Math. Proc. Cambridge Phil. Soc.}\ }\textbf
  {\bibinfo {volume} {85}},\ \bibinfo {pages} {465} (\bibinfo {year}
  {1979})}\BibitemShut {NoStop}%
\bibitem [{\citenamefont {Calabi}(1979)}]{ASENS_1979_4_12_2_269_0}%
  \BibitemOpen
  \bibfield  {author} {\bibinfo {author} {\bibfnamefont {E.}~\bibnamefont
  {Calabi}},\ }\href {\doibase 10.24033/asens.1367} {\bibfield  {journal}
  {\bibinfo  {journal} {Annales scientifiques de l'\'Ecole Normale
  Sup\'erieure}\ }\textbf {\bibinfo {volume} {4e s{\'e}rie, 12}},\ \bibinfo
  {pages} {269} (\bibinfo {year} {1979})}\BibitemShut {NoStop}%
\bibitem [{\citenamefont {Townsend}(1977)}]{Townsend:1977xw}%
  \BibitemOpen
  \bibfield  {author} {\bibinfo {author} {\bibfnamefont {P.~K.}\ \bibnamefont
  {Townsend}},\ }\href {\doibase 10.1103/PhysRevD.15.2795} {\bibfield
  {journal} {\bibinfo  {journal} {Phys. Rev. D}\ }\textbf {\bibinfo {volume}
  {15}},\ \bibinfo {pages} {2795} (\bibinfo {year} {1977})}\BibitemShut
  {NoStop}%
\bibitem [{\citenamefont {MacDowell}\ and\ \citenamefont
  {Mansouri}(1977)}]{MacDowell:1977jt}%
  \BibitemOpen
  \bibfield  {author} {\bibinfo {author} {\bibfnamefont {S.~W.}\ \bibnamefont
  {MacDowell}}\ and\ \bibinfo {author} {\bibfnamefont {F.}~\bibnamefont
  {Mansouri}},\ }\href {\doibase 10.1103/PhysRevLett.38.739} {\bibfield
  {journal} {\bibinfo  {journal} {Phys. Rev. Lett.}\ }\textbf {\bibinfo
  {volume} {38}},\ \bibinfo {pages} {739} (\bibinfo {year} {1977})},\ \bibinfo
  {note} {[Erratum: Phys. Rev. Lett. 38, 1376 (1977)]}\BibitemShut {NoStop}%
\bibitem [{\citenamefont {Andrianopoli}\ and\ \citenamefont
  {D'Auria}(2014)}]{Andrianopoli:2014aqa}%
  \BibitemOpen
  \bibfield  {author} {\bibinfo {author} {\bibfnamefont {L.}~\bibnamefont
  {Andrianopoli}}\ and\ \bibinfo {author} {\bibfnamefont {R.}~\bibnamefont
  {D'Auria}},\ }\href {\doibase 10.1007/JHEP08(2014)012} {\bibfield  {journal}
  {\bibinfo  {journal} {JHEP}\ }\textbf {\bibinfo {volume} {08}},\ \bibinfo
  {pages} {012} (\bibinfo {year} {2014})}\BibitemShut {NoStop}%
\bibitem [{\citenamefont {Andrianopoli}\ \emph {et~al.}(2021)\citenamefont
  {Andrianopoli}, \citenamefont {Cerchiai}, \citenamefont {Matrecano},
  \citenamefont {Miskovic}, \citenamefont {Noris}, \citenamefont {Olea},
  \citenamefont {Ravera},\ and\ \citenamefont
  {Trigiante}}]{Andrianopoli:2020zbl}%
  \BibitemOpen
  \bibfield  {author} {\bibinfo {author} {\bibfnamefont {L.}~\bibnamefont
  {Andrianopoli}}, \bibinfo {author} {\bibfnamefont {B.~L.}\ \bibnamefont
  {Cerchiai}}, \bibinfo {author} {\bibfnamefont {R.}~\bibnamefont {Matrecano}},
  \bibinfo {author} {\bibfnamefont {O.}~\bibnamefont {Miskovic}}, \bibinfo
  {author} {\bibfnamefont {R.}~\bibnamefont {Noris}}, \bibinfo {author}
  {\bibfnamefont {R.}~\bibnamefont {Olea}}, \bibinfo {author} {\bibfnamefont
  {L.}~\bibnamefont {Ravera}}, \ and\ \bibinfo {author} {\bibfnamefont
  {M.}~\bibnamefont {Trigiante}},\ }\href {\doibase 10.1007/JHEP02(2021)141}
  {\bibfield  {journal} {\bibinfo  {journal} {JHEP}\ }\textbf {\bibinfo
  {volume} {02}},\ \bibinfo {pages} {141} (\bibinfo {year} {2021})}\BibitemShut
  {NoStop}%
\bibitem [{\citenamefont {Lukierski}\ and\ \citenamefont
  {Nowicki}(1983)}]{Lukierski:1983jg}%
  \BibitemOpen
  \bibfield  {author} {\bibinfo {author} {\bibfnamefont {J.}~\bibnamefont
  {Lukierski}}\ and\ \bibinfo {author} {\bibfnamefont {A.}~\bibnamefont
  {Nowicki}},\ }\href {\doibase 10.1016/0370-2693(83)91626-X} {\bibfield
  {journal} {\bibinfo  {journal} {Phys. Lett. B}\ }\textbf {\bibinfo {volume}
  {127}},\ \bibinfo {pages} {40} (\bibinfo {year} {1983})}\BibitemShut
  {NoStop}%
\bibitem [{\citenamefont {Lukierski}\ and\ \citenamefont
  {Nowicki}(1986)}]{Lukierski:1983qc}%
  \BibitemOpen
  \bibfield  {author} {\bibinfo {author} {\bibfnamefont {J.}~\bibnamefont
  {Lukierski}}\ and\ \bibinfo {author} {\bibfnamefont {A.}~\bibnamefont
  {Nowicki}},\ }\href {\doibase 10.1016/0003-4916(86)90054-0} {\bibfield
  {journal} {\bibinfo  {journal} {Annals Phys.}\ }\textbf {\bibinfo {volume}
  {166}},\ \bibinfo {pages} {164} (\bibinfo {year} {1986})}\BibitemShut
  {NoStop}%
\bibitem [{\citenamefont {Lukierski}\ and\ \citenamefont
  {Zakrzewski}(1987)}]{Lukierski:1986jw}%
  \BibitemOpen
  \bibfield  {author} {\bibinfo {author} {\bibfnamefont {J.}~\bibnamefont
  {Lukierski}}\ and\ \bibinfo {author} {\bibfnamefont {W.~J.}\ \bibnamefont
  {Zakrzewski}},\ }\href {\doibase 10.1016/0370-2693(87)91277-9} {\bibfield
  {journal} {\bibinfo  {journal} {Phys. Lett. B}\ }\textbf {\bibinfo {volume}
  {189}},\ \bibinfo {pages} {99} (\bibinfo {year} {1987})}\BibitemShut
  {NoStop}%
\bibitem [{\citenamefont {Miskovic}\ and\ \citenamefont
  {Olea}(2009)}]{Miskovic:2009bm}%
  \BibitemOpen
  \bibfield  {author} {\bibinfo {author} {\bibfnamefont {O.}~\bibnamefont
  {Miskovic}}\ and\ \bibinfo {author} {\bibfnamefont {R.}~\bibnamefont
  {Olea}},\ }\href {\doibase 10.1103/PhysRevD.79.124020} {\bibfield  {journal}
  {\bibinfo  {journal} {Phys. Rev. D}\ }\textbf {\bibinfo {volume} {79}},\
  \bibinfo {pages} {124020} (\bibinfo {year} {2009})}\BibitemShut {NoStop}%
\bibitem [{\citenamefont {Araneda}\ \emph {et~al.}(2016)\citenamefont
  {Araneda}, \citenamefont {Aros}, \citenamefont {Miskovic},\ and\
  \citenamefont {Olea}}]{Araneda:2016iiy}%
  \BibitemOpen
  \bibfield  {author} {\bibinfo {author} {\bibfnamefont {R.}~\bibnamefont
  {Araneda}}, \bibinfo {author} {\bibfnamefont {R.}~\bibnamefont {Aros}},
  \bibinfo {author} {\bibfnamefont {O.}~\bibnamefont {Miskovic}}, \ and\
  \bibinfo {author} {\bibfnamefont {R.}~\bibnamefont {Olea}},\ }\href {\doibase
  10.1103/PhysRevD.93.084022} {\bibfield  {journal} {\bibinfo  {journal} {Phys.
  Rev. D}\ }\textbf {\bibinfo {volume} {93}},\ \bibinfo {pages} {084022}
  (\bibinfo {year} {2016})}\BibitemShut {NoStop}%
\bibitem [{\citenamefont {Lynden-Bell}\ and\ \citenamefont
  {Nouri-Zonoz}(1998)}]{Lynden-Bell:1996dpw}%
  \BibitemOpen
  \bibfield  {author} {\bibinfo {author} {\bibfnamefont {D.}~\bibnamefont
  {Lynden-Bell}}\ and\ \bibinfo {author} {\bibfnamefont {M.}~\bibnamefont
  {Nouri-Zonoz}},\ }\href {\doibase 10.1103/RevModPhys.70.427} {\bibfield
  {journal} {\bibinfo  {journal} {Rev. Mod. Phys.}\ }\textbf {\bibinfo {volume}
  {70}},\ \bibinfo {pages} {427} (\bibinfo {year} {1998})}\BibitemShut
  {NoStop}%
\bibitem [{\citenamefont {Misner}(1963)}]{Misner:1963fr}%
  \BibitemOpen
  \bibfield  {author} {\bibinfo {author} {\bibfnamefont {C.~W.}\ \bibnamefont
  {Misner}},\ }\href {\doibase 10.1063/1.1704019} {\bibfield  {journal}
  {\bibinfo  {journal} {J. Math. Phys.}\ }\textbf {\bibinfo {volume} {4}},\
  \bibinfo {pages} {924} (\bibinfo {year} {1963})}\BibitemShut {NoStop}%
\bibitem [{\citenamefont {Hawking}(1977)}]{Hawking:1976jb}%
  \BibitemOpen
  \bibfield  {author} {\bibinfo {author} {\bibfnamefont {S.~W.}\ \bibnamefont
  {Hawking}},\ }\href {\doibase 10.1016/0375-9601(77)90386-3} {\bibfield
  {journal} {\bibinfo  {journal} {Phys. Lett. A}\ }\textbf {\bibinfo {volume}
  {60}},\ \bibinfo {pages} {81} (\bibinfo {year} {1977})}\BibitemShut {NoStop}%
\bibitem [{\citenamefont {Eguchi}\ \emph {et~al.}(1978)\citenamefont {Eguchi},
  \citenamefont {Gilkey},\ and\ \citenamefont {Hanson}}]{Eguchi:1977iu}%
  \BibitemOpen
  \bibfield  {author} {\bibinfo {author} {\bibfnamefont {T.}~\bibnamefont
  {Eguchi}}, \bibinfo {author} {\bibfnamefont {P.~B.}\ \bibnamefont {Gilkey}},
  \ and\ \bibinfo {author} {\bibfnamefont {A.~J.}\ \bibnamefont {Hanson}},\
  }\href {\doibase 10.1103/PhysRevD.17.423} {\bibfield  {journal} {\bibinfo
  {journal} {Phys. Rev. D}\ }\textbf {\bibinfo {volume} {17}},\ \bibinfo
  {pages} {423} (\bibinfo {year} {1978})}\BibitemShut {NoStop}%
\bibitem [{\citenamefont {Gibbons}\ and\ \citenamefont
  {Hawking}(1979)}]{Gibbons:1979xm}%
  \BibitemOpen
  \bibfield  {author} {\bibinfo {author} {\bibfnamefont {G.~W.}\ \bibnamefont
  {Gibbons}}\ and\ \bibinfo {author} {\bibfnamefont {S.~W.}\ \bibnamefont
  {Hawking}},\ }\href {\doibase 10.1007/BF01197189} {\bibfield  {journal}
  {\bibinfo  {journal} {Commun. Math. Phys.}\ }\textbf {\bibinfo {volume}
  {66}},\ \bibinfo {pages} {291} (\bibinfo {year} {1979})}\BibitemShut
  {NoStop}%
\bibitem [{\citenamefont {Cl\'ement}\ \emph {et~al.}(2015)\citenamefont
  {Cl\'ement}, \citenamefont {Gal'tsov},\ and\ \citenamefont
  {Guenouche}}]{Clement:2015cxa}%
  \BibitemOpen
  \bibfield  {author} {\bibinfo {author} {\bibfnamefont {G.}~\bibnamefont
  {Cl\'ement}}, \bibinfo {author} {\bibfnamefont {D.}~\bibnamefont {Gal'tsov}},
  \ and\ \bibinfo {author} {\bibfnamefont {M.}~\bibnamefont {Guenouche}},\
  }\href {\doibase 10.1016/j.physletb.2015.09.074} {\bibfield  {journal}
  {\bibinfo  {journal} {Phys. Lett. B}\ }\textbf {\bibinfo {volume} {750}},\
  \bibinfo {pages} {591} (\bibinfo {year} {2015})},\ \Eprint
  {http://arxiv.org/abs/1508.07622} {1508.07622} \BibitemShut {NoStop}%
\bibitem [{\citenamefont {Hennigar}\ \emph {et~al.}(2019)\citenamefont
  {Hennigar}, \citenamefont {Kubiz\v{n}\'ak},\ and\ \citenamefont
  {Mann}}]{Hennigar:2019ive}%
  \BibitemOpen
  \bibfield  {author} {\bibinfo {author} {\bibfnamefont {R.~A.}\ \bibnamefont
  {Hennigar}}, \bibinfo {author} {\bibfnamefont {D.}~\bibnamefont
  {Kubiz\v{n}\'ak}}, \ and\ \bibinfo {author} {\bibfnamefont {R.~B.}\
  \bibnamefont {Mann}},\ }\href {\doibase 10.1103/PhysRevD.100.064055}
  {\bibfield  {journal} {\bibinfo  {journal} {Phys. Rev. D}\ }\textbf {\bibinfo
  {volume} {100}},\ \bibinfo {pages} {064055} (\bibinfo {year}
  {2019})}\BibitemShut {NoStop}%
\bibitem [{\citenamefont {Bordo}\ \emph
  {et~al.}(2019{\natexlab{a}})\citenamefont {Bordo}, \citenamefont {Gray},\
  and\ \citenamefont {Kubiz\v{n}\'ak}}]{Bordo:2019slw}%
  \BibitemOpen
  \bibfield  {author} {\bibinfo {author} {\bibfnamefont {A.~B.}\ \bibnamefont
  {Bordo}}, \bibinfo {author} {\bibfnamefont {F.}~\bibnamefont {Gray}}, \ and\
  \bibinfo {author} {\bibfnamefont {D.}~\bibnamefont {Kubiz\v{n}\'ak}},\ }\href
  {\doibase 10.1007/JHEP07(2019)119} {\bibfield  {journal} {\bibinfo  {journal}
  {JHEP}\ }\textbf {\bibinfo {volume} {07}},\ \bibinfo {pages} {119} (\bibinfo
  {year} {2019}{\natexlab{a}})}\BibitemShut {NoStop}%
\bibitem [{\citenamefont {Bordo}\ \emph
  {et~al.}(2019{\natexlab{b}})\citenamefont {Bordo}, \citenamefont {Gray},
  \citenamefont {Hennigar},\ and\ \citenamefont
  {Kubiz\v{n}\'ak}}]{Bordo:2019tyh}%
  \BibitemOpen
  \bibfield  {author} {\bibinfo {author} {\bibfnamefont {A.~B.}\ \bibnamefont
  {Bordo}}, \bibinfo {author} {\bibfnamefont {F.}~\bibnamefont {Gray}},
  \bibinfo {author} {\bibfnamefont {R.~A.}\ \bibnamefont {Hennigar}}, \ and\
  \bibinfo {author} {\bibfnamefont {D.}~\bibnamefont {Kubiz\v{n}\'ak}},\ }\href
  {\doibase 10.1088/1361-6382/ab3d4d} {\bibfield  {journal} {\bibinfo
  {journal} {Class. Quant. Grav.}\ }\textbf {\bibinfo {volume} {36}},\ \bibinfo
  {pages} {194001} (\bibinfo {year} {2019}{\natexlab{b}})}\BibitemShut
  {NoStop}%
\bibitem [{\citenamefont {Ballon~Bordo}\ \emph {et~al.}(2019)\citenamefont
  {Ballon~Bordo}, \citenamefont {Gray}, \citenamefont {Hennigar},\ and\
  \citenamefont {Kubiz\v{n}\'ak}}]{BallonBordo:2019vrn}%
  \BibitemOpen
  \bibfield  {author} {\bibinfo {author} {\bibfnamefont {A.}~\bibnamefont
  {Ballon~Bordo}}, \bibinfo {author} {\bibfnamefont {F.}~\bibnamefont {Gray}},
  \bibinfo {author} {\bibfnamefont {R.~A.}\ \bibnamefont {Hennigar}}, \ and\
  \bibinfo {author} {\bibfnamefont {D.}~\bibnamefont {Kubiz\v{n}\'ak}},\ }\href
  {\doibase 10.1016/j.physletb.2019.134972} {\bibfield  {journal} {\bibinfo
  {journal} {Phys. Lett. B}\ }\textbf {\bibinfo {volume} {798}},\ \bibinfo
  {pages} {134972} (\bibinfo {year} {2019})}\BibitemShut {NoStop}%
\bibitem [{\citenamefont {Durka}(2022)}]{Durka:2019ajz}%
  \BibitemOpen
  \bibfield  {author} {\bibinfo {author} {\bibfnamefont {R.}~\bibnamefont
  {Durka}},\ }\href {\doibase 10.1142/S0218271822500213} {\bibfield  {journal}
  {\bibinfo  {journal} {Int. J. Mod. Phys. D}\ }\textbf {\bibinfo {volume}
  {31}},\ \bibinfo {pages} {2250021} (\bibinfo {year} {2022})}\BibitemShut
  {NoStop}%
\bibitem [{\citenamefont {Wu}\ and\ \citenamefont {Wu}(2019)}]{Wu:2019pzr}%
  \BibitemOpen
  \bibfield  {author} {\bibinfo {author} {\bibfnamefont {S.-Q.}\ \bibnamefont
  {Wu}}\ and\ \bibinfo {author} {\bibfnamefont {D.}~\bibnamefont {Wu}},\ }\href
  {\doibase 10.1103/PhysRevD.100.101501} {\bibfield  {journal} {\bibinfo
  {journal} {Phys. Rev. D}\ }\textbf {\bibinfo {volume} {100}},\ \bibinfo
  {pages} {101501} (\bibinfo {year} {2019})}\BibitemShut {NoStop}%
\bibitem [{\citenamefont {Chen}\ and\ \citenamefont
  {Jiang}(2019)}]{Chen:2019uhp}%
  \BibitemOpen
  \bibfield  {author} {\bibinfo {author} {\bibfnamefont {Z.}~\bibnamefont
  {Chen}}\ and\ \bibinfo {author} {\bibfnamefont {J.}~\bibnamefont {Jiang}},\
  }\href {\doibase 10.1103/PhysRevD.100.104016} {\bibfield  {journal} {\bibinfo
   {journal} {Phys. Rev. D}\ }\textbf {\bibinfo {volume} {100}},\ \bibinfo
  {pages} {104016} (\bibinfo {year} {2019})}\BibitemShut {NoStop}%
\bibitem [{\citenamefont {Ballon~Bordo}\ \emph {et~al.}(2020)\citenamefont
  {Ballon~Bordo}, \citenamefont {Gray},\ and\ \citenamefont
  {Kubiz\v{n}\'ak}}]{BallonBordo:2020mcs}%
  \BibitemOpen
  \bibfield  {author} {\bibinfo {author} {\bibfnamefont {A.}~\bibnamefont
  {Ballon~Bordo}}, \bibinfo {author} {\bibfnamefont {F.}~\bibnamefont {Gray}},
  \ and\ \bibinfo {author} {\bibfnamefont {D.}~\bibnamefont {Kubiz\v{n}\'ak}},\
  }\href {\doibase 10.1007/JHEP05(2020)084} {\bibfield  {journal} {\bibinfo
  {journal} {JHEP}\ }\textbf {\bibinfo {volume} {05}},\ \bibinfo {pages} {084}
  (\bibinfo {year} {2020})}\BibitemShut {NoStop}%
\bibitem [{\citenamefont {Riegert}(1984)}]{Riegert:1984zz}%
  \BibitemOpen
  \bibfield  {author} {\bibinfo {author} {\bibfnamefont {R.~J.}\ \bibnamefont
  {Riegert}},\ }\href {\doibase 10.1103/PhysRevLett.53.315} {\bibfield
  {journal} {\bibinfo  {journal} {Phys. Rev. Lett.}\ }\textbf {\bibinfo
  {volume} {53}},\ \bibinfo {pages} {315} (\bibinfo {year} {1984})}\BibitemShut
  {NoStop}%
\bibitem [{\citenamefont {Lu}\ \emph {et~al.}(2012)\citenamefont {Lu},
  \citenamefont {Pang}, \citenamefont {Pope},\ and\ \citenamefont
  {Vazquez-Poritz}}]{Lu:2012xu}%
  \BibitemOpen
  \bibfield  {author} {\bibinfo {author} {\bibfnamefont {H.}~\bibnamefont
  {Lu}}, \bibinfo {author} {\bibfnamefont {Y.}~\bibnamefont {Pang}}, \bibinfo
  {author} {\bibfnamefont {C.~N.}\ \bibnamefont {Pope}}, \ and\ \bibinfo
  {author} {\bibfnamefont {J.~F.}\ \bibnamefont {Vazquez-Poritz}},\ }\href
  {\doibase 10.1103/PhysRevD.86.044011} {\bibfield  {journal} {\bibinfo
  {journal} {Phys. Rev. D}\ }\textbf {\bibinfo {volume} {86}},\ \bibinfo
  {pages} {044011} (\bibinfo {year} {2012})}\BibitemShut {NoStop}%
\bibitem [{\citenamefont {Liu}\ \emph {et~al.}(2013)\citenamefont {Liu},
  \citenamefont {L\"u}, \citenamefont {Pope},\ and\ \citenamefont
  {V\'azquez-Poritz}}]{Liu:2013fna}%
  \BibitemOpen
  \bibfield  {author} {\bibinfo {author} {\bibfnamefont {H.-S.}\ \bibnamefont
  {Liu}}, \bibinfo {author} {\bibfnamefont {H.}~\bibnamefont {L\"u}}, \bibinfo
  {author} {\bibfnamefont {C.~N.}\ \bibnamefont {Pope}}, \ and\ \bibinfo
  {author} {\bibfnamefont {J.~F.}\ \bibnamefont {V\'azquez-Poritz}},\ }\href
  {\doibase 10.1088/0264-9381/30/16/165015} {\bibfield  {journal} {\bibinfo
  {journal} {Class. Quant. Grav.}\ }\textbf {\bibinfo {volume} {30}},\ \bibinfo
  {pages} {165015} (\bibinfo {year} {2013})}\BibitemShut {NoStop}%
\bibitem [{\citenamefont {Corral}\ \emph {et~al.}(2021)\citenamefont {Corral},
  \citenamefont {Giribet},\ and\ \citenamefont {Olea}}]{Corral:2021xsu}%
  \BibitemOpen
  \bibfield  {author} {\bibinfo {author} {\bibfnamefont {C.}~\bibnamefont
  {Corral}}, \bibinfo {author} {\bibfnamefont {G.}~\bibnamefont {Giribet}}, \
  and\ \bibinfo {author} {\bibfnamefont {R.}~\bibnamefont {Olea}},\ }\href
  {\doibase 10.1103/PhysRevD.104.064026} {\bibfield  {journal} {\bibinfo
  {journal} {Phys. Rev. D}\ }\textbf {\bibinfo {volume} {104}},\ \bibinfo
  {pages} {064026} (\bibinfo {year} {2021})}\BibitemShut {NoStop}%
\bibitem [{\citenamefont {Maldacena}(2011)}]{Maldacena:2011mk}%
  \BibitemOpen
  \bibfield  {author} {\bibinfo {author} {\bibfnamefont {J.}~\bibnamefont
  {Maldacena}},\ }\href@noop {} {\  (\bibinfo {year} {2011})},\ \Eprint
  {http://arxiv.org/abs/1105.5632} {arXiv:1105.5632 [hep-th]} \BibitemShut
  {NoStop}%
\bibitem [{\citenamefont {Anastasiou}\ and\ \citenamefont
  {Olea}(2016)}]{Anastasiou:2016jix}%
  \BibitemOpen
  \bibfield  {author} {\bibinfo {author} {\bibfnamefont {G.}~\bibnamefont
  {Anastasiou}}\ and\ \bibinfo {author} {\bibfnamefont {R.}~\bibnamefont
  {Olea}},\ }\href {\doibase 10.1103/PhysRevD.94.086008} {\bibfield  {journal}
  {\bibinfo  {journal} {Phys. Rev. D}\ }\textbf {\bibinfo {volume} {94}},\
  \bibinfo {pages} {086008} (\bibinfo {year} {2016})}\BibitemShut {NoStop}%
\bibitem [{\citenamefont {Anastasiou}\ \emph {et~al.}(2021)\citenamefont
  {Anastasiou}, \citenamefont {Araya},\ and\ \citenamefont
  {Olea}}]{Anastasiou:2020mik}%
  \BibitemOpen
  \bibfield  {author} {\bibinfo {author} {\bibfnamefont {G.}~\bibnamefont
  {Anastasiou}}, \bibinfo {author} {\bibfnamefont {I.~J.}\ \bibnamefont
  {Araya}}, \ and\ \bibinfo {author} {\bibfnamefont {R.}~\bibnamefont {Olea}},\
  }\href {\doibase 10.1007/JHEP01(2021)134} {\bibfield  {journal} {\bibinfo
  {journal} {JHEP}\ }\textbf {\bibinfo {volume} {01}},\ \bibinfo {pages} {134}
  (\bibinfo {year} {2021})}\BibitemShut {NoStop}%
\bibitem [{\citenamefont {Alvarez}\ \emph
  {et~al.}(2014{\natexlab{b}})\citenamefont {Alvarez}, \citenamefont {Pais},
  \citenamefont {Rodr\'\i{}guez}, \citenamefont {Salgado-Rebolledo},\ and\
  \citenamefont {Zanelli}}]{Alvarez:2014uda}%
  \BibitemOpen
  \bibfield  {author} {\bibinfo {author} {\bibfnamefont {P.~D.}\ \bibnamefont
  {Alvarez}}, \bibinfo {author} {\bibfnamefont {P.}~\bibnamefont {Pais}},
  \bibinfo {author} {\bibfnamefont {E.}~\bibnamefont {Rodr\'\i{}guez}},
  \bibinfo {author} {\bibfnamefont {P.}~\bibnamefont {Salgado-Rebolledo}}, \
  and\ \bibinfo {author} {\bibfnamefont {J.}~\bibnamefont {Zanelli}},\ }\href
  {\doibase 10.1016/j.physletb.2014.09.032} {\bibfield  {journal} {\bibinfo
  {journal} {Phys. Lett. B}\ }\textbf {\bibinfo {volume} {738}},\ \bibinfo
  {pages} {134} (\bibinfo {year} {2014}{\natexlab{b}})}\BibitemShut {NoStop}%
\bibitem [{\citenamefont {Zanelli}(2005)}]{Zanelli:2005sa}%
  \BibitemOpen
  \bibfield  {author} {\bibinfo {author} {\bibfnamefont {J.}~\bibnamefont
  {Zanelli}},\ }in\ \href@noop {} {\emph {\bibinfo {booktitle} {{7th Mexican
  Workshop on Particles and Fields}}}}\ (\bibinfo {year} {2005})\ \Eprint
  {http://arxiv.org/abs/hep-th/0502193} {arXiv:hep-th/0502193} \BibitemShut
  {NoStop}%
\bibitem [{\citenamefont {Hassaine}\ and\ \citenamefont
  {Zanelli}(2016)}]{Hassaine:2016amq}%
  \BibitemOpen
  \bibfield  {author} {\bibinfo {author} {\bibfnamefont {M.}~\bibnamefont
  {Hassaine}}\ and\ \bibinfo {author} {\bibfnamefont {J.}~\bibnamefont
  {Zanelli}},\ }\href {\doibase 10.1142/9863} {\emph {\bibinfo {title}
  {{Chern-Simons (super)gravity}}}},\ \bibinfo {edition} {2016th}\ ed.,\
  \bibinfo {series} {100 years of general relativity}, Vol.~\bibinfo {volume}
  {2}\ (\bibinfo  {publisher} {World Scientific},\ \bibinfo {address}
  {Hackensack},\ \bibinfo {year} {2016})\BibitemShut {NoStop}%
\bibitem [{\citenamefont {Alvarez}\ \emph
  {et~al.}(2023{\natexlab{b}})\citenamefont {Alvarez}, \citenamefont {Corral},\
  and\ \citenamefont {Zanelli}}]{Alvarez:2022wcj}%
  \BibitemOpen
  \bibfield  {author} {\bibinfo {author} {\bibfnamefont {P.~D.}\ \bibnamefont
  {Alvarez}}, \bibinfo {author} {\bibfnamefont {C.}~\bibnamefont {Corral}}, \
  and\ \bibinfo {author} {\bibfnamefont {J.}~\bibnamefont {Zanelli}},\ }\href
  {\doibase 10.1007/JHEP01(2023)009} {\bibfield  {journal} {\bibinfo  {journal}
  {JHEP}\ }\textbf {\bibinfo {volume} {01}},\ \bibinfo {pages} {009} (\bibinfo
  {year} {2023}{\natexlab{b}})}\BibitemShut {NoStop}%
\bibitem [{\citenamefont {Page}(1978)}]{Page:1978hdy}%
  \BibitemOpen
  \bibfield  {author} {\bibinfo {author} {\bibfnamefont {D.~N.}\ \bibnamefont
  {Page}},\ }\href {\doibase 10.1016/0370-2693(78)90016-3} {\bibfield
  {journal} {\bibinfo  {journal} {Phys. Lett. B}\ }\textbf {\bibinfo {volume}
  {78}},\ \bibinfo {pages} {249} (\bibinfo {year} {1978})}\BibitemShut
  {NoStop}%
\bibitem [{\citenamefont {Pedersen}(1985)}]{Pedersen_1985}%
  \BibitemOpen
  \bibfield  {author} {\bibinfo {author} {\bibfnamefont {H.}~\bibnamefont
  {Pedersen}},\ }\href {\doibase 10.1088/0264-9381/2/4/022} {\bibfield
  {journal} {\bibinfo  {journal} {Classical and Quantum Gravity}\ }\textbf
  {\bibinfo {volume} {2}},\ \bibinfo {pages} {579} (\bibinfo {year}
  {1985})}\BibitemShut {NoStop}%
\bibitem [{\citenamefont {Nitta}\ and\ \citenamefont
  {Uzawa}(2021)}]{Nitta:2020pzo}%
  \BibitemOpen
  \bibfield  {author} {\bibinfo {author} {\bibfnamefont {M.}~\bibnamefont
  {Nitta}}\ and\ \bibinfo {author} {\bibfnamefont {K.}~\bibnamefont {Uzawa}},\
  }\href {\doibase 10.1140/epjc/s10052-021-09289-0} {\bibfield  {journal}
  {\bibinfo  {journal} {Eur. Phys. J. C}\ }\textbf {\bibinfo {volume} {81}},\
  \bibinfo {pages} {513} (\bibinfo {year} {2021})}\BibitemShut {NoStop}%
\bibitem [{\citenamefont {Clarkson}\ and\ \citenamefont
  {Mann}(2006{\natexlab{a}})}]{Clarkson:2005qx}%
  \BibitemOpen
  \bibfield  {author} {\bibinfo {author} {\bibfnamefont {R.}~\bibnamefont
  {Clarkson}}\ and\ \bibinfo {author} {\bibfnamefont {R.~B.}\ \bibnamefont
  {Mann}},\ }\href {\doibase 10.1088/0264-9381/23/5/005} {\bibfield  {journal}
  {\bibinfo  {journal} {Class. Quant. Grav.}\ }\textbf {\bibinfo {volume}
  {23}},\ \bibinfo {pages} {1507} (\bibinfo {year}
  {2006}{\natexlab{a}})}\BibitemShut {NoStop}%
\bibitem [{\citenamefont {Clarkson}\ and\ \citenamefont
  {Mann}(2006{\natexlab{b}})}]{Clarkson:2006zk}%
  \BibitemOpen
  \bibfield  {author} {\bibinfo {author} {\bibfnamefont {R.}~\bibnamefont
  {Clarkson}}\ and\ \bibinfo {author} {\bibfnamefont {R.~B.}\ \bibnamefont
  {Mann}},\ }\href {\doibase 10.1103/PhysRevLett.96.051104} {\bibfield
  {journal} {\bibinfo  {journal} {Phys. Rev. Lett.}\ }\textbf {\bibinfo
  {volume} {96}},\ \bibinfo {pages} {051104} (\bibinfo {year}
  {2006}{\natexlab{b}})}\BibitemShut {NoStop}%
\bibitem [{\citenamefont {Ishihara}\ \emph {et~al.}(2006)\citenamefont
  {Ishihara}, \citenamefont {Kimura}, \citenamefont {Matsuno},\ and\
  \citenamefont {Tomizawa}}]{Ishihara:2006pb}%
  \BibitemOpen
  \bibfield  {author} {\bibinfo {author} {\bibfnamefont {H.}~\bibnamefont
  {Ishihara}}, \bibinfo {author} {\bibfnamefont {M.}~\bibnamefont {Kimura}},
  \bibinfo {author} {\bibfnamefont {K.}~\bibnamefont {Matsuno}}, \ and\
  \bibinfo {author} {\bibfnamefont {S.}~\bibnamefont {Tomizawa}},\ }\href
  {\doibase 10.1103/PhysRevD.74.047501} {\bibfield  {journal} {\bibinfo
  {journal} {Phys. Rev. D}\ }\textbf {\bibinfo {volume} {74}},\ \bibinfo
  {pages} {047501} (\bibinfo {year} {2006})}\BibitemShut {NoStop}%
\bibitem [{\citenamefont {Chng}\ \emph {et~al.}(2006)\citenamefont {Chng},
  \citenamefont {Mann},\ and\ \citenamefont {Stelea}}]{Chng:2006gh}%
  \BibitemOpen
  \bibfield  {author} {\bibinfo {author} {\bibfnamefont {B.}~\bibnamefont
  {Chng}}, \bibinfo {author} {\bibfnamefont {R.~B.}\ \bibnamefont {Mann}}, \
  and\ \bibinfo {author} {\bibfnamefont {C.}~\bibnamefont {Stelea}},\ }\href
  {\doibase 10.1103/PhysRevD.74.084031} {\bibfield  {journal} {\bibinfo
  {journal} {Phys. Rev. D}\ }\textbf {\bibinfo {volume} {74}},\ \bibinfo
  {pages} {084031} (\bibinfo {year} {2006})}\BibitemShut {NoStop}%
\bibitem [{\citenamefont {Wong}\ and\ \citenamefont
  {Mann}(2012{\natexlab{a}})}]{Wong:2011aa}%
  \BibitemOpen
  \bibfield  {author} {\bibinfo {author} {\bibfnamefont {A.~W.~C.}\
  \bibnamefont {Wong}}\ and\ \bibinfo {author} {\bibfnamefont {R.~B.}\
  \bibnamefont {Mann}},\ }\href {\doibase 10.1103/PhysRevD.85.046010}
  {\bibfield  {journal} {\bibinfo  {journal} {Phys. Rev. D}\ }\textbf {\bibinfo
  {volume} {85}},\ \bibinfo {pages} {046010} (\bibinfo {year}
  {2012}{\natexlab{a}})}\BibitemShut {NoStop}%
\bibitem [{\citenamefont {Wong}\ and\ \citenamefont
  {Mann}(2012{\natexlab{b}})}]{Wong:2012ne}%
  \BibitemOpen
  \bibfield  {author} {\bibinfo {author} {\bibfnamefont {A.~W.~C.}\
  \bibnamefont {Wong}}\ and\ \bibinfo {author} {\bibfnamefont {R.~B.}\
  \bibnamefont {Mann}},\ }\href {\doibase 10.1103/PhysRevD.86.124002}
  {\bibfield  {journal} {\bibinfo  {journal} {Phys. Rev. D}\ }\textbf {\bibinfo
  {volume} {86}},\ \bibinfo {pages} {124002} (\bibinfo {year}
  {2012}{\natexlab{b}})}\BibitemShut {NoStop}%
\bibitem [{\citenamefont {Hendi}\ \emph {et~al.}(2012)\citenamefont {Hendi},
  \citenamefont {Mann}, \citenamefont {Riazi},\ and\ \citenamefont
  {Eslam~Panah}}]{Hendi:2012zg}%
  \BibitemOpen
  \bibfield  {author} {\bibinfo {author} {\bibfnamefont {S.~H.}\ \bibnamefont
  {Hendi}}, \bibinfo {author} {\bibfnamefont {R.~B.}\ \bibnamefont {Mann}},
  \bibinfo {author} {\bibfnamefont {N.}~\bibnamefont {Riazi}}, \ and\ \bibinfo
  {author} {\bibfnamefont {B.}~\bibnamefont {Eslam~Panah}},\ }\href {\doibase
  10.1103/PhysRevD.86.104034} {\bibfield  {journal} {\bibinfo  {journal} {Phys.
  Rev. D}\ }\textbf {\bibinfo {volume} {86}},\ \bibinfo {pages} {104034}
  (\bibinfo {year} {2012})}\BibitemShut {NoStop}%
\bibitem [{\citenamefont {Durgut}\ \emph {et~al.}(2023)\citenamefont {Durgut},
  \citenamefont {Hennigar}, \citenamefont {Kunduri},\ and\ \citenamefont
  {Mann}}]{Durgut:2022xzw}%
  \BibitemOpen
  \bibfield  {author} {\bibinfo {author} {\bibfnamefont {T.}~\bibnamefont
  {Durgut}}, \bibinfo {author} {\bibfnamefont {R.~A.}\ \bibnamefont
  {Hennigar}}, \bibinfo {author} {\bibfnamefont {H.~K.}\ \bibnamefont
  {Kunduri}}, \ and\ \bibinfo {author} {\bibfnamefont {R.~B.}\ \bibnamefont
  {Mann}},\ }\href {\doibase 10.1007/JHEP03(2023)114} {\bibfield  {journal}
  {\bibinfo  {journal} {JHEP}\ }\textbf {\bibinfo {volume} {03}},\ \bibinfo
  {pages} {114} (\bibinfo {year} {2023})}\BibitemShut {NoStop}%
\bibitem [{\citenamefont {Cembranos}\ and\ \citenamefont
  {Valcarcel}(2017)}]{Cembranos:2016gdt}%
  \BibitemOpen
  \bibfield  {author} {\bibinfo {author} {\bibfnamefont {J.~A.~R.}\
  \bibnamefont {Cembranos}}\ and\ \bibinfo {author} {\bibfnamefont {J.~G.}\
  \bibnamefont {Valcarcel}},\ }\href {\doibase 10.1088/1475-7516/2017/01/014}
  {\bibfield  {journal} {\bibinfo  {journal} {JCAP}\ }\textbf {\bibinfo
  {volume} {01}},\ \bibinfo {pages} {014} (\bibinfo {year} {2017})}\BibitemShut
  {NoStop}%
\bibitem [{\citenamefont {Cembranos}\ and\ \citenamefont
  {Gigante~Valcarcel}(2018)}]{Cembranos:2017pcs}%
  \BibitemOpen
  \bibfield  {author} {\bibinfo {author} {\bibfnamefont {J.~A.~R.}\
  \bibnamefont {Cembranos}}\ and\ \bibinfo {author} {\bibfnamefont
  {J.}~\bibnamefont {Gigante~Valcarcel}},\ }\href {\doibase
  10.1016/j.physletb.2018.01.081} {\bibfield  {journal} {\bibinfo  {journal}
  {Phys. Lett. B}\ }\textbf {\bibinfo {volume} {779}},\ \bibinfo {pages} {143}
  (\bibinfo {year} {2018})}\BibitemShut {NoStop}%
\bibitem [{\citenamefont {Obukhov}(2020)}]{Obukhov:2020hlp}%
  \BibitemOpen
  \bibfield  {author} {\bibinfo {author} {\bibfnamefont {Y.~N.}\ \bibnamefont
  {Obukhov}},\ }\href {\doibase 10.1103/PhysRevD.102.104059} {\bibfield
  {journal} {\bibinfo  {journal} {Phys. Rev. D}\ }\textbf {\bibinfo {volume}
  {102}},\ \bibinfo {pages} {104059} (\bibinfo {year} {2020})}\BibitemShut
  {NoStop}%
\bibitem [{\citenamefont {Xiao}(2004)}]{Xiao_2004}%
  \BibitemOpen
  \bibfield  {author} {\bibinfo {author} {\bibfnamefont {Z.}~\bibnamefont
  {Xiao}},\ }\href {\doibase 10.1088/0253-6102/42/2/235} {\bibfield  {journal}
  {\bibinfo  {journal} {Communications in Theoretical Physics}\ }\textbf
  {\bibinfo {volume} {42}},\ \bibinfo {pages} {235} (\bibinfo {year}
  {2004})}\BibitemShut {NoStop}%
\bibitem [{\citenamefont {Chen}\ and\ \citenamefont
  {Zhang}(2021)}]{Chen:2020org}%
  \BibitemOpen
  \bibfield  {author} {\bibinfo {author} {\bibfnamefont {J.}~\bibnamefont
  {Chen}}\ and\ \bibinfo {author} {\bibfnamefont {X.}~\bibnamefont {Zhang}},\
  }\href {\doibase 10.1016/j.geomphys.2020.104010} {\bibfield  {journal}
  {\bibinfo  {journal} {J. Geom. Phys.}\ }\textbf {\bibinfo {volume} {161}},\
  \bibinfo {pages} {104010} (\bibinfo {year} {2021})}\BibitemShut {NoStop}%
\bibitem [{\citenamefont {Yang}\ and\ \citenamefont
  {Zhang}(2023)}]{Yang:2023snt}%
  \BibitemOpen
  \bibfield  {author} {\bibinfo {author} {\bibfnamefont {Y.}~\bibnamefont
  {Yang}}\ and\ \bibinfo {author} {\bibfnamefont {X.}~\bibnamefont {Zhang}},\
  }\href {\doibase 10.1140/epjc/s10052-023-11762-x} {\bibfield  {journal}
  {\bibinfo  {journal} {Eur. Phys. J. C}\ }\textbf {\bibinfo {volume} {83}},\
  \bibinfo {pages} {574} (\bibinfo {year} {2023})}\BibitemShut {NoStop}%
\bibitem [{\citenamefont {Johnson}(2014{\natexlab{a}})}]{Johnson:2014xza}%
  \BibitemOpen
  \bibfield  {author} {\bibinfo {author} {\bibfnamefont {C.~V.}\ \bibnamefont
  {Johnson}},\ }\href {\doibase 10.1088/0264-9381/31/23/235003} {\bibfield
  {journal} {\bibinfo  {journal} {Class. Quant. Grav.}\ }\textbf {\bibinfo
  {volume} {31}},\ \bibinfo {pages} {235003} (\bibinfo {year}
  {2014}{\natexlab{a}})}\BibitemShut {NoStop}%
\bibitem [{\citenamefont {Johnson}(2014{\natexlab{b}})}]{Johnson:2014pwa}%
  \BibitemOpen
  \bibfield  {author} {\bibinfo {author} {\bibfnamefont {C.~V.}\ \bibnamefont
  {Johnson}},\ }\href {\doibase 10.1088/0264-9381/31/22/225005} {\bibfield
  {journal} {\bibinfo  {journal} {Class. Quant. Grav.}\ }\textbf {\bibinfo
  {volume} {31}},\ \bibinfo {pages} {225005} (\bibinfo {year}
  {2014}{\natexlab{b}})}\BibitemShut {NoStop}%
\bibitem [{\citenamefont {Atiyah}\ \emph {et~al.}(1973)\citenamefont {Atiyah},
  \citenamefont {Patodi},\ and\ \citenamefont {Singer}}]{APS}%
  \BibitemOpen
  \bibfield  {author} {\bibinfo {author} {\bibfnamefont {M.~F.}\ \bibnamefont
  {Atiyah}}, \bibinfo {author} {\bibfnamefont {V.~K.}\ \bibnamefont {Patodi}},
  \ and\ \bibinfo {author} {\bibfnamefont {I.~M.}\ \bibnamefont {Singer}},\
  }\href {\doibase https://doi.org/10.1112/blms/5.2.229} {\bibfield  {journal}
  {\bibinfo  {journal} {Bulletin of the London Mathematical Society}\ }\textbf
  {\bibinfo {volume} {5}},\ \bibinfo {pages} {229} (\bibinfo {year}
  {1973})}\BibitemShut {NoStop}%
\bibitem [{\citenamefont {Hitchin}(1974)}]{HITCHIN19741}%
  \BibitemOpen
  \bibfield  {author} {\bibinfo {author} {\bibfnamefont {N.}~\bibnamefont
  {Hitchin}},\ }\href {\doibase https://doi.org/10.1016/0001-8708(74)90021-8}
  {\bibfield  {journal} {\bibinfo  {journal} {Advances in Mathematics}\
  }\textbf {\bibinfo {volume} {14}},\ \bibinfo {pages} {1} (\bibinfo {year}
  {1974})}\BibitemShut {NoStop}%
\bibitem [{\citenamefont {Hojman}\ \emph {et~al.}(1980)\citenamefont {Hojman},
  \citenamefont {Mukku},\ and\ \citenamefont {Sayed}}]{Hojman:1980kv}%
  \BibitemOpen
  \bibfield  {author} {\bibinfo {author} {\bibfnamefont {R.}~\bibnamefont
  {Hojman}}, \bibinfo {author} {\bibfnamefont {C.}~\bibnamefont {Mukku}}, \
  and\ \bibinfo {author} {\bibfnamefont {W.~A.}\ \bibnamefont {Sayed}},\ }\href
  {\doibase 10.1103/PhysRevD.22.1915} {\bibfield  {journal} {\bibinfo
  {journal} {Phys. Rev. D}\ }\textbf {\bibinfo {volume} {22}},\ \bibinfo
  {pages} {1915} (\bibinfo {year} {1980})}\BibitemShut {NoStop}%
\bibitem [{\citenamefont {Nelson}(1980)}]{Nelson:1980ph}%
  \BibitemOpen
  \bibfield  {author} {\bibinfo {author} {\bibfnamefont {P.~C.}\ \bibnamefont
  {Nelson}},\ }\href {\doibase 10.1016/0375-9601(80)90348-5} {\bibfield
  {journal} {\bibinfo  {journal} {Phys. Lett. A}\ }\textbf {\bibinfo {volume}
  {79}},\ \bibinfo {pages} {285} (\bibinfo {year} {1980})}\BibitemShut
  {NoStop}%
\bibitem [{\citenamefont {Nieh}\ and\ \citenamefont {Yan}(1982)}]{Nieh:1981ww}%
  \BibitemOpen
  \bibfield  {author} {\bibinfo {author} {\bibfnamefont {H.~T.}\ \bibnamefont
  {Nieh}}\ and\ \bibinfo {author} {\bibfnamefont {M.~L.}\ \bibnamefont {Yan}},\
  }\href {\doibase 10.1063/1.525379} {\bibfield  {journal} {\bibinfo  {journal}
  {J. Math. Phys.}\ }\textbf {\bibinfo {volume} {23}},\ \bibinfo {pages} {373}
  (\bibinfo {year} {1982})}\BibitemShut {NoStop}%
\bibitem [{\citenamefont {Chandia}\ and\ \citenamefont
  {Zanelli}(1997)}]{Chandia:1997hu}%
  \BibitemOpen
  \bibfield  {author} {\bibinfo {author} {\bibfnamefont {O.}~\bibnamefont
  {Chandia}}\ and\ \bibinfo {author} {\bibfnamefont {J.}~\bibnamefont
  {Zanelli}},\ }\href {\doibase 10.1103/PhysRevD.55.7580} {\bibfield  {journal}
  {\bibinfo  {journal} {Phys. Rev. D}\ }\textbf {\bibinfo {volume} {55}},\
  \bibinfo {pages} {7580} (\bibinfo {year} {1997})}\BibitemShut {NoStop}%
\end{thebibliography}%

\end{document}